\documentclass[fleqn,usenatbib]{mnras}

\usepackage{newtxtext,newtxmath}

\usepackage[T1]{fontenc}

\DeclareRobustCommand{\VAN}[3]{#2}
\let\VANthebibliography\thebibliography
\def\thebibliography{\DeclareRobustCommand{\VAN}[3]{##3}\VANthebibliography}

\usepackage{graphicx}	
\usepackage{amsmath}	

\usepackage{tikz,xcolor,hyperref}

\definecolor{lime}{HTML}{A6CE39}
\DeclareRobustCommand{\orcidicon}{%
\begin{tikzpicture}
\draw[lime, fill=lime] (0,0)
circle [radius=0.16]
node[white] {{\fontfamily{qag}\selectfont \tiny ID}};
\draw[white, fill=white] (-0.0625,0.095)
circle [radius=0.007];
\end{tikzpicture}
\hspace{-4mm}
}

\foreach \x in {A, ..., Z}{%
\expandafter\xdef\csname orcid\x\endcsname{\noexpand\href{https://orcid.org/\csname orcidauthor\x\endcsname}{\noexpand\orcidicon}}
}

\title[Stellar Populations of Ring Galaxies]{Integral Field Spectroscopy of Collisional Ring Galaxies I: Stellar Populations Analysis}

\author[Chow-Mart\'inez et al.]{
M. Chow-Mart\'inez$^{1,2}$\orcidA{}\,\,\thanks{E-mail: marcel.chow@unan.edu.ni},
A. Robleto-Orús$^{1,2}$\orcidB{}\,\,,
Y.D. Mayya$^{3}$\orcidC{}\,\,,
J.P. Torres-Papaqui$^{4}$\orcidD{}\,\,,
\newauthor
R.A. Ortega-Minakata$^{5}$\orcidE{}\,\,,
D.F. Castro-Hidalgo$^{6}$\orcidF{}\,\,,
C.A. Caretta$^{4}$\orcidG{}\,\,,
J.J. Trejo-Alonso$^{7}$\orcidH{}\,\,,
\newauthor
A. Morales-Vargas$^{8}$\orcidI{}\,\,,
R. Garc\'ia-Benito$^{9}$\orcidJ{}\,\,,
H.E. Jácamo-Delgado$^{10}$\orcidK{}\,\,,
M. Gudiño$^{11}$\orcidM{}
\\
$^{1}$Instituto de Geolog\'ia y Geof\'isica Benjamin Linder y Héroes de Bocay (IGG-BLyHB), Universidad Nacional Aut\'onoma de Nicaragua, Managua (UNAN-Managua),\\ C.P. 663, Managua, Nicaragua\\
$^{2}$Centro de Investigaci\'on de Astrof\'isica y Ciencias Espaciales (CIACE), Universidad Nacional Aut\'onoma de Nicaragua, Managua (UNAN-Managua),\\ C.P. 663, Managua, Nicaragua\\
$^{3}$Instituto Nacional de Astrof\'isica, \'Optica y Electr\'onica, Luis E. Erro, Tonantzintla, 72840 Puebla, Mexico\\
$^{4}$Departamento de Astronom\'ia, Universidad de Guanajuato, Callejón de Jalisco S/N, Col. Valenciana, C.P. 36023, Guanajuato, Gto., Mexico\\
$^{5}$Instituto de Radioastronom\'ia y Astrof\'isica (IRyA), UNAM, Apartado Postal 72-3, Morelia, 58089 Michoac\'an, Mexico\\
$^{6}$Departamento de Ciencias Exactas, Centro Universitario de la Costa, Universidad de Guadalajara, Puerto Vallarta, Jalisco C.P. 48280, Mexico\\
$^{7}$Facultad de Ingenier\'ia, Universidad Aut\'onoma de Quer\'etaro, Cerro de las Campanas s/n, 76010, Santiago de Quer\'etaro, Qro., Mexico\\
$^{8}$Centro de Astronom\'ia (CITEVA), Universidad de Antofagasta, Avenida U. de Antofagasta 02800, Antofagasta, Chile\\
$^{9}$ Instituto de Astrof\'isica de Andaluc\'ia – CSIC, Glorieta de la Astronom\'ia s/n, 18008 Granada, Spain\\
$^{10}$Departamento de F\'isica, \'Area de Ciencias B\'asicas e Ingenier\'ia, Universidad Nacional Autónoma de Nicaragua, Managua (UNAN-Managua), C.P. 663,\\  Managua, Nicaragua\\
$^{11}$Instituto de Astronom\'ia, Universidad Nacional Aut\'onoma de M\'exico,  AP 70-264, 04510, Ciudad de M\'exico, Mexico
}

\date{Accepted XXX. Received YYY; in original form ZZZ}

\pubyear{202-}

\begin{document}
\label{firstpage}
\pagerange{\pageref{firstpage}--\pageref{lastpage}}
\maketitle

\begin{abstract}
    Collisional ring galaxies are produced by the collision of a disk galaxy with a compact galaxy plunging through the disk, forming a ring-shaped expanding density wave, triggering star formation at its wake. The wave expansion is expected to produce negative stellar age gradients in radial profiles of post-collision stellar populations. Integral field spectroscopy combined with stellar population synthesis allows us to spatially resolve the stellar populations, to separate the post-collision and pre-collision components, and to produce the radial profiles. We analyse three candidate galaxies: Arp~143, NGC~2793, and VII~Zw~466. Observations were performed with the Calar Alto 3.5~m~telescope using the PMAS/PPak spectrophotometer. NGC 2793 presents a positive stellar age gradient, dismissing the collision hypothesis. For Arp~143 and VII~Zw~466, we found negative stellar age gradients for the youngest stellar populations, up to the ring radii, consistent with the collision hypothesis. We estimated that the collisions occurred $\sim$300~Myr and $\sim$100~Myr (expansion velocities of 33~$\pm$~10 km s$^{-1}$ and 108~$\pm$~26 km s$^{-1}$), respectively, before the density waves reached the observed ring radii. A spatially resolved analysis of the specific star formation histories (sSFH), reveals an expected star formation enhancement following the collision. The sSFH also allowed to identify the most probable intruder galaxy for VII~Zw~466. We report new redshifts for its group members. Finally, radial profiles of light contributions from pre-collisional and post-collisional stars show that the density wave dragged old pre-collisional stars along, as predicted by simulations.
\end{abstract}



\section{Introduction}

Ring galaxies are one of the less common morphological types among peculiar galaxies. They are believed to be produced by an infrequent kind of collision \citep{burbidge1959,athanassoula1985}: a compact galaxy plunging almost perpendicular through the equatorial plane of a disk galaxy. In this scenario the ring is formed by star-forming regions induced by the passage of a density wave produced by the collision.

This hypothesis was supported by early N-body simulations \citep[e.g.,][]{lynds1976,theys1977,mapelli2008}. Later simulations including stellar and gas dynamics, reproduced different ring morphologies and other associated structures, such as the spokes seen in the famous Cartwheel ring galaxy \citep{hernquist1993, Struck-Marcell1993}. \citealt{gerber1996} simulations considered the formation and propagation of the ring based on a Milky Way like galaxy to reproduce the features of the Cartwheel. They found an expansion velocity of the ring $\sim$140 $\mathrm{km \ s^{-1}}$ and rotation speeds in the ring up to 450~$\mathrm{km \ s^{-1}}$. However, this predicted expansion velocities are much higher compared to the 13-30~$\mathrm{km \ s^{-1}}$ obtained observationally  \citep{higdon1996,amram1998}. These velocities give Cartwheel kinematical ages between 150 and 400~Myr. 

An observational consequence of star formation in an expanding wave is the formation of a colour gradient in the disk galaxy as new generation of stars is formed at successively outer parts as the density wave passes through. The youngest and brightest blue stars of this new generation dominate the observed light of the star-forming ring in most bands. As the density wave, and the corresponding star-forming ring, advances farther from the centre, it leaves in its wake stellar remnants of massive stars and low-mass stars that mix with the the stellar population of  the pre-collisional disk. As a result, a colour gradient should be observed with bluer colours (proxy of younger stellar ages) at the position of the ring, and redder colours (proxy of older stellar ages) closer to the impact centre. Most of the observational attempts to test this scenario have focused on the Cartwheel galaxy, given its proximity and relatively high angular size \citep{marcum1992, korchagin2001, vorobyov2001}.

As for other ring galaxies, the compilation by \citet{romano2008} is the only systematic study to date testing the collision hypothesis for a set of ring galaxy candidates, taken from the \citet{appleton1987} sample. \citet{romano2008} measured the H$\mathrm{\alpha}$ intensity and colour radial profiles for fifteen ring galaxy candidates, showing that at least nine of them present radial colour gradients consistent with the propagation of the star formation wave. However they found that the observed colour gradient has an important contribution from the intrinsic colour gradient existing in the disks of galaxies.  

The last decade has seen drastic advances in both computational and observational fronts that are impacting all fields of astronomy. The most notable advancement in the simulation of ring galaxies is the one by \citet{renaud2018}, who carried out hydrodynamical simulations of Cartwheel-kind of interactions using adaptive mesh refinement code to couple the galactic scale dynamics with star-formation reaching spatial scales of 6~pc. According to these more realistic simulations, the newly formed stars are dragged by the wave before falling back into the central regions at later stages. In this expanding material wave scenario, a wide range of ages are expected in the ring regions.

\citet{zaragoza2022} carried out such a study of the Cartwheel galaxy using the Multi Unit Spectroscopic Explorer (MUSE) data. They found that the radial gradient of the present-day gas-phase metallicities supports the predictions of \citet{renaud2018} scenario of an expanding material wave. More recently, \citet{ditrani2024} derived star formation histories (SFH) from the same MUSE data. They concluded that the observed spectral flux of the ring is dominated mostly by the youngest, luminous post-collisional stars. On the other hand, \citealt{mayya2024}, using the UV images of the Cartwheel galaxy, reported that the star-forming ring consists of stellar populations formed over the last 150~Myr. They also found evidence for the presence of an age gradient of the populations formed in the wake of the expanding wave as predicted by the classical density wave models. These observations suggest that the correct picture is somewhere in between with some of the stars formed by the expanding wave remaining at the location of their formation while some stars are dragged by the wave.

The most notable advance on the observational front is the Integral Field Spectroscopy \citep[IFS, e.g.][]{dezeeuw2002,sanchez2012,bundy2015}, 
which allows spatially resolved analysis of the spectra over the entire galaxy using stellar population synthesis models \citep[e.g.][]{conroy2013,vazdekis2016,maraston2020} to disentangle the current star formation rate \citep[e.g.][]{catalan2015,catalan2017,ellison2018,morales2020} as well as the complete star formation histories \citep[e.g.][]{ibarra2016,lopez2019,ditrani2024}. This technique allows to study the spatial distribution of the stellar populations and their properties such as their age, metallicity and kinematics. In this work, we present IFS observations of three ring galaxies---selected from the \citet{romano2008} sample---in order to test the collision hypothesis by searching for age gradients: Arp\,143, NGC\,2793 and VII\,Zw\,466. While \cite{romano2008} used colour gradients as proxies for stellar age gradients, the IFS observations combined with stellar population synthesis models allows us to infer the presence of stellar populations in the disks from both the post and pre-collisional star formation through the detection of their characteristic spectral features. We found that both Arp 143 and VII Zw 466 are consistent with what is expected for collisional ring galaxies, while NGC~2793 appears to be an ordinary dwarf galaxy. In order to explain how we got to this conclusion, we show here the analysis on these three galaxies.
 
The paper is organized as follows: the observations and the reduction process are presented in section \ref{sec:obs}. Section \ref{sec:method} shows the methodology and analysis of the data. The results are described in section \ref{sec:res}. Finally, section \ref{sec:con} discusses the conclusions. Throughout the paper, a standard cosmology, with a Hubble-Lemaitre constant of $H_0=70$ km s$^{-1}$, $\Omega_\Lambda = 0.7$ and $\Omega_\textrm{M} = 0.3$ is assumed.

\section{Observations and data Reduction}
\label{sec:obs}

\begin{table*}
        \centering
            \caption{General description of the sample of observed galaxies. For VII~Zw~466 we observed two offset but overlapped fields to analyse the companion galaxies in the group, searching for the possible colliding companion. Coordinates and B$_r$ magnitudes (in Vega system) are taken from the \protect\cite{appleton1987} catalogue. These coordinates are close to but not exactly at the centre of the image, nor at the centre of the galaxy. Ring radii were taken from \protect\cite{romano2008} and correspond to the major semi-axis derived from H$\alpha$ emission ellipses fit.}
        \label{tab:obs_gal}
        \begin{tabular}{lccccccccc} %
        \hline
        Galaxy  & RA       & Dec               & Redshift & Redshift              & Distance & B$_{r}$   & Ring             &  Exp. Time & Mean  \\
                & (h m s)    &(\degr\,\,\arcmin\,\,\arcsec) &          & Reference             & (Mpc)    & (mag)     & radius (\arcsec) &       (h)      &  Airmass        \\
        \hline
       Arp~143     & 07 46 52  &  +39 00 36   &  0.01295 & \citealt{galbany2018} & 59       & 13.6      & 36               & 2.25           & 1.04     \\
       NGC~2793    & 09 16 47  &  +34 25 47   &  0.00563 & \citealt{springob2005}& 28       & 14.2      & 20               & 2.25           & 1.05     \\
       VII Zw  466 & 12 31 56  &  +66 24 41   &  0.04833 & \citealt{corwin1994}  & 215      & 14.6      & 11               & 4.50           & 1.15     \\
        \hline
        \end{tabular}
\end{table*}

The observations were carried out with the 3.5 m telescope at the Calar Alto Observatory (Almer\'ia, Spain), using the PMAS/PPak spectrophotometer, across two nights (2016 March 28$^{\rm th}$ and 30$^{\rm th}$). We observed eight galaxies from the \citet{romano2008} sample. Three of them are presented in this paper (see Table~\ref{tab:obs_gal}).

During the observations, there were acceptable weather conditions with a mean seeing FWHM of 1.5~arcsec. We used the V500 (R~$\sim$850) and V1200 (R~$\sim$1650) gratings for the three objects, with integration times of $900$~s and $1800$~s, respectively, with three exposures per field per grating. This allowed us to adopt the three-pointing dithering scheme used by the CALIFA survey (The Calar Alto Legacy Integral Field Area Survey, \citealt{sanchez2012}). This scheme reaches a 100\% filling factor across the entire field~of~view (FoV) by combining three slightly offset pointings of the telescope, compensating for the spaces between the instrument optical fibres. In addition to science observations, sky-flats and standard stars spectra were acquired for the reduction process.

The reduction was performed using the pipeline described in \citet{husemann2013}, \citet{garciabenito2015} and \citet{sanchez2016}, following the CALIFA reduction scheme: the raw observation pass through a semi-automatic reduction, including bias subtraction, cosmic rays removal, arc lamp calibration, wavelength solution and flux calibration.

The three dithering exposures of each field for each grating were combined into a single data cube. Then, the V1200 data cubes were spectrally combined with the V500 into a combined (COMB) data cube, by averaging the flux in the overlapping wavelength region and resampling the V1200 spectral range to V500 resolution. This strategy compensates for internal vignetting in the blue range (3745 – 4240~\AA) of the V500 grating \citep[see][for details]{sanchez2012}. The resulting spectral sampling is 2.0~\AA, 0.7~\AA ~ and 2.0~\AA, for the V500, V1200 and COMB data cubes, respectively. For this work, we used only the resulting COMB data cubes with a combined spectral range from 3745~\AA ~ to 7501~\AA. The nominal resolution of these combined spectra is FWHM $\sim$ 6.5~\AA, actually the same as V500.

\section{Analysis methods}
\label{sec:method}

\subsection{Stellar population synthesis}
\label{sec:SPS}
We performed the stellar population synthesis using the \textsf{STARLIGHT} code \citep{cidfernandes2005}. \textsf{STARLIGHT} fits each spectrum to a combination of simple stellar populations (SSPs), by estimating the percentage of each SSP contributes to the spectrum starlight flux, to the ``present-day'' stellar mass, and to the initial stellar mass. 

We used a selection of SSPs based on the work of \citet{mateus2006}, comprising 150 SSPs covering 25 ages (logarithmically spaced from 1~Myr to 18~Gyr) and 6 metallicities: Z $=$ 0.005, 0.02, 0.2, 0.4, 1.0 and 2.5 Z$_\odot$. The chosen age and metallicity ranges include values expected for both pre- and post-collision star formation events. These SSPs correspond to the 2016 version of the \citet{bruzual2003} stellar population models\footnote{\url{http://www.bruzual.org/bc03/Updated_version_2016/}}, at a resolution with a FWHM of 2.3~\AA.
While we follow the general approach of \citet{mateus2006}, we introduce a modification by adopting the MILES stellar library \citep{sanchezblazquez2006, vazdekis2010, falconbarroso2011}. We retain the original Padova stellar evolutionary tracks \citep{bertelli1994} and the \citet{chabrier2003} Initial Mass Function (IMF), and apply the reddening law of \citet{cardelli1989} with $R_V = 3.1$. This recipe has satisfactorily been used by \citet{morales2020} for PMAS/PPak data.

\begin{figure*}
 \includegraphics[width=0.94\textwidth]{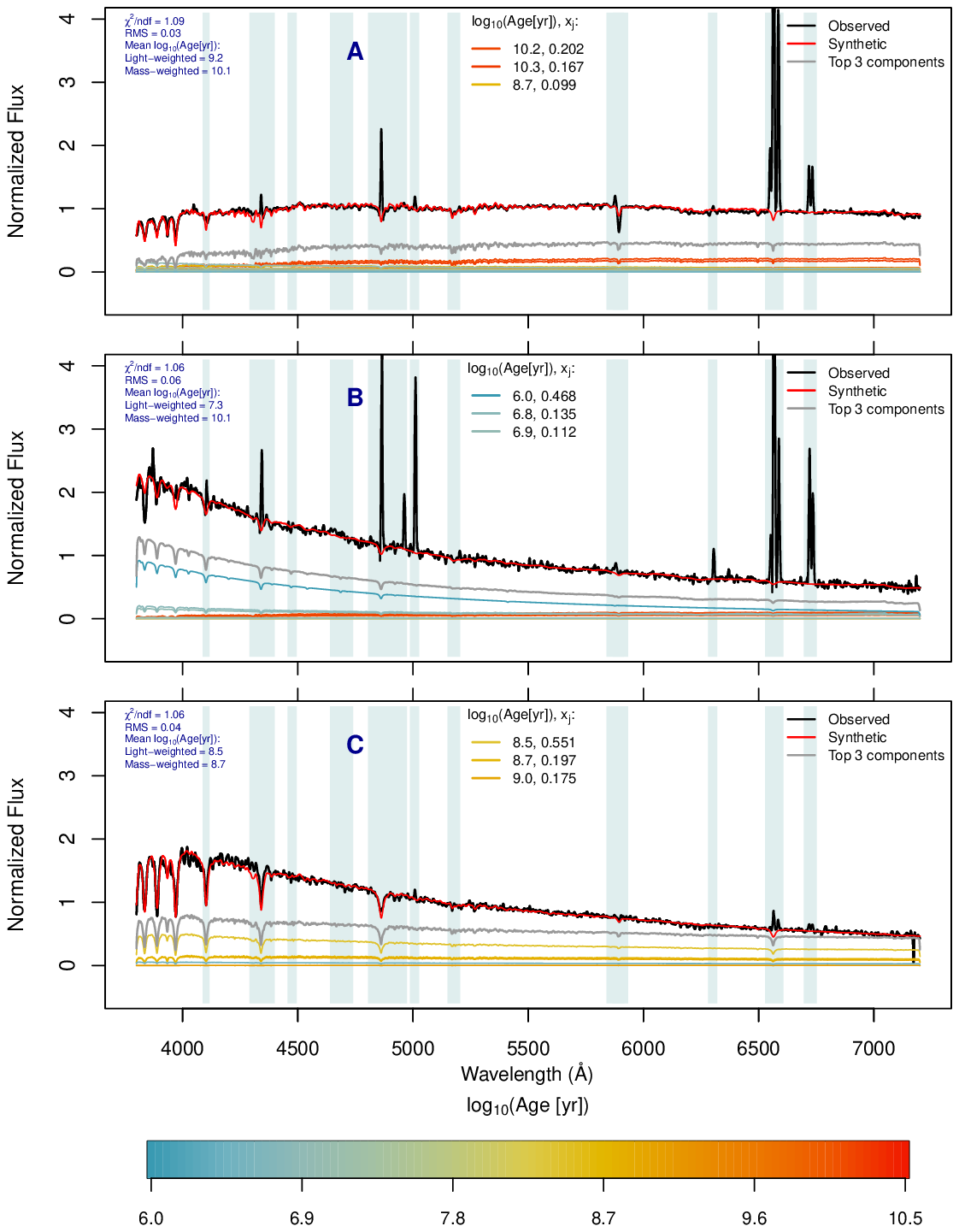}
        \caption{Representative spectral fits for three spaxels in Arp~143. The locations of these spaxels, labelled as A, B, and C, are indicated in Figure~\ref{fig:age_map}, top-left panel. The flux of each spectrum is normalized at $\lambda = 5100$~\AA. The observed spectrum is plotted with a solid-black line and the synthetic spectrum with the best-fit solution is plotted with a red-dashed line. The individual SSP spectra contributing to the solution are plotted as solid coloured lines. Each of these lines represents an SSP spectrum of a given age, integrated over the six metallicities, and scaled by its corresponding population vector component, $x_j$. The colour of each line corresponds to the age of that stellar population, as indicated by the colour bar at the bottom. In each panel, the central inset legend lists the ages of the three SSPs that most significantly contribute to the synthetic spectrum (the Top 3 components). The solid gray line represents the sum of these three SSPs. The shaded regions indicate the spectral intervals that were masked out during the \textsf{STARLIGHT} fit to exclude emission lines. The goodness-of-fit parameters, $\chi^2/$ndf and the RMS, are shown in the top-left corner, along with the light-weighted and mass-weighted mean stellar ages for each spaxel, computed using Equations~\ref{eq:xmean_age} and \ref{eq:mmean_age}}.
 \label{fig:spec_grid}
\end{figure*}

\begin{figure*}
 \includegraphics[width=0.99\textwidth]{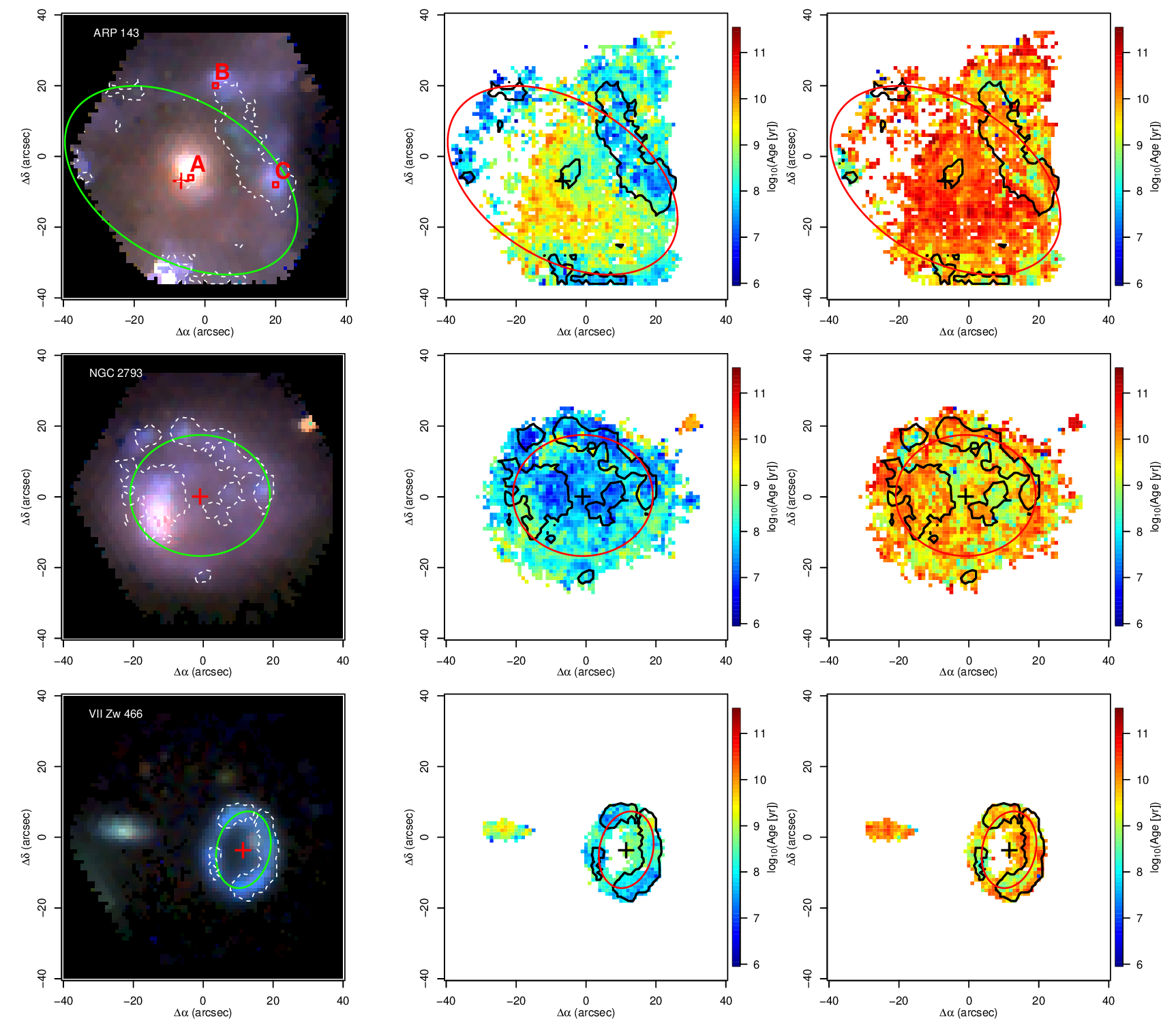}
        \caption{Left panels: RGB images of the observed galaxies produced by extracting the BVR Johnson synthetic bands from the IFS data cubes. Central panels: Light-Weighted Mean Age maps ($ \langle \log{t_\star} \rangle_L$). Right panels: Mass-Weighted Mean Age maps ($ \langle \log{t_\star} \rangle_M$). Contours in all maps represent H$\alpha$ surface brightness at 0.1 $L_\odot$ kpc$^{-2}$ obtained from spaxels with  S/N > 3 for H$\alpha$. Ellipses were taken from the \protect\cite{romano2008} fit over the H$\alpha$ band, with the crosses marking the centres of those ellipses. The relative coordinates of these maps are in relation to the IFS image centre (see coordinates in Table \ref{tab:obs_gal}). North is up and East is to the left. For Arp~143 (top-left panel), the red labels A, B, and C correspond to the position of the spaxels presented in Figure~\ref{fig:spec_grid}.
        }
 \label{fig:age_map}
\end{figure*}

\begin{figure}
 \includegraphics[width=0.95\columnwidth]{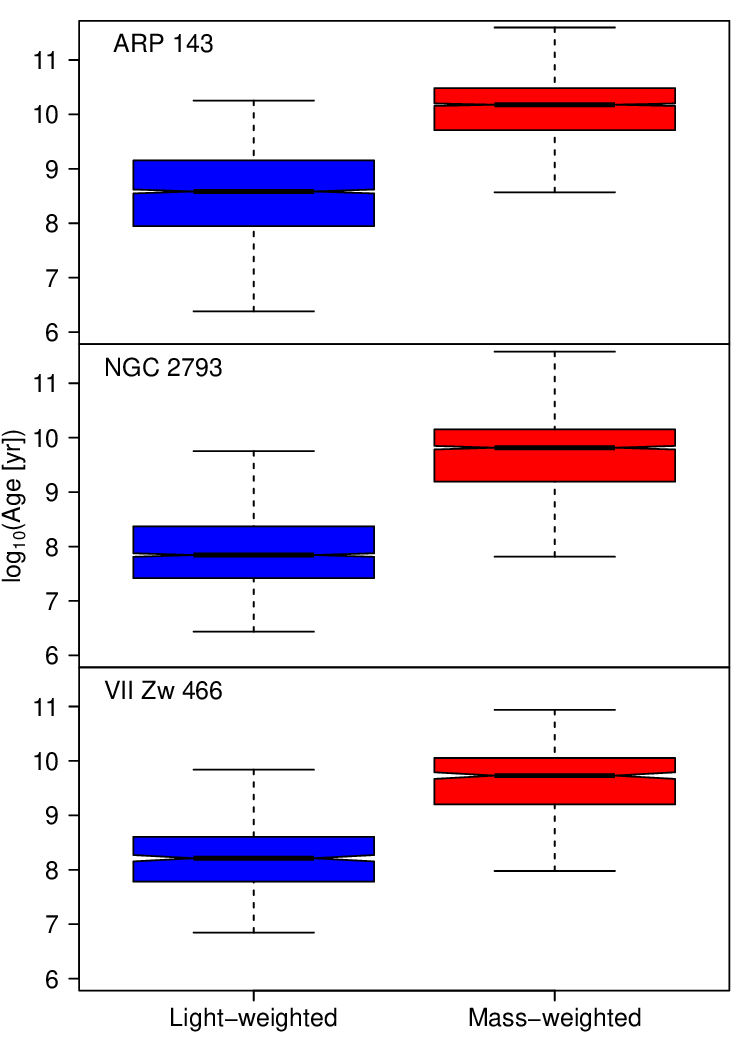}
        \caption{Box plots of the distribution of light-weighted ($\langle \log{t_\star} \rangle_L$)  and mass-weighted ($\langle \log{t_\star} \rangle_M$) ages for the three galaxies of our sample. As for any classical box plot, the box contains data points between the first and third quartiles, and the whiskers represent the distance to the farthest data points within 1.5 of the interquartile range (IQR). Bold lines mark the median, while the notches represent the confidence interval around that median. The S/N > 10 criterion was applied beforehand. For VII~Zw~466, only spaxels on the ring galaxy were included.}
 \label{fig:age_boxplot}
\end{figure}

The spectra extracted from the COMB data cubes were prepared for the \textsf{STARLIGHT} fitting process by shifting the spectra to the galaxy rest frame, resampling to a wavelength interval of $\Delta \lambda = 1.0$~\AA\ and masking out the emission lines, following the STARLIGHT manual\footnote{Available at \url{http://www.starlight.ufsc.br/}}. Errors were propagated using a covariance matrix accounting for both the fraction of flux within each new $\Delta \lambda$ interval and the quadrature sum of the uncertainties from the original spectra (see \citealt{carnall2017} for details).

\textsf{STARLIGHT} then uses a Markov Chain Monte Carlo (MCMC) method to fit a linear combination of the SSPs to the observed spectrum. The resulting fitted continuum is convolved by a Gaussian kernel whose FWHM is determined by the MCMC. That FWHM is similar (but not exactly equal) to the instrumental FWHM of the observed spectrum combined in quadrature with the FWHM of the SSP spectra used for the linear combination.


The fitting was restricted to the spectral range from 3800~\AA\ to 7200~\AA\ in order to avoid edge effects. The correction for galactic extinction was applied by using the dust maps produced by \citealt{schlegel1998}. We only accepted \textsf{STARLIGHT} solutions for spaxels with S/N >10 in the stellar continuum, computed in the window from 5075~\AA\ to 5125~\AA. Representative examples of the spectral fits are presented in Figure\ref{fig:spec_grid}.

We calculated the light-weighted mean age\footnote{Weighted by the light contribution of the $j$-th population to the wavelength range of the spectrum.} $ \langle \log{t_\star} \rangle_L$ and mass-weighted mean age  $ \langle \log{t_\star} \rangle_M$, as defined by \cite{asari2007}:

\begin{equation}
        \label{eq:xmean_age}
        \langle \log{t_\star} \rangle_L = \sum^{N_\star}_{j=1} x_j\log_{10}{t_{\star,j}}
\end{equation}

\noindent
and

\begin{equation}
        \label{eq:mmean_age}
        \langle \log{t_\star} \rangle_M = \sum^{N_\star}_{j=1} \mu_j \log_{10}{t_{\star,j}}
\end{equation}
\noindent
where $x_j$ and $\mu_j$ are the starlight flux fraction and stellar mass fraction corresponding to the $j$-th SSP, respectively, while $t_{\star,j}$ is the age of the $j$-th SSP and $N_\star$ is the total number of SSPs. 

Equations \ref{eq:xmean_age} and \ref{eq:mmean_age} tend to highlight different stellar populations: the light-weighted mean ages tend to yield younger populations, as short-lived massive stars (O, B, and A types) dominate the optical flux when present, despite being less numerous than low-mass stars.

On the other hand, the mass-weighted mean ages tend to yield long-lived low-mass stars (such as G, K and M stars), because of their large mass-to-light ratio as well as because of their typically large numbers in giant galaxies. The core of these tends is in the statistical nature of the definitions of $\langle \log{t_\star} \rangle_L$ and $\langle \log{t_\star} \rangle_M$. For a deeper discussion on this topic, see \cite{cidfernandes2003} and \cite{cidfernandes2004}.
 
\textsf{STARLIGHT} estimates several properties of the stellar populations as output for each spectrum, including mass, stellar extinction, velocity shift and velocity dispersion. The \textsf{STARLIGHT} solution also produces a synthetic spectrum for the stellar continuum, that is the sum of the flux contribution of all the SSPs to each wavelength (including dust extinction), being a good approximation to the observed stellar continuum without the emission features.

\subsection{Emission line analysis}
The fitted synthetic stellar continuum from each spaxel was subtracted from the observed spectrum to obtain a ``pure emission'' spectrum. We fitted single-Gaussian models to the profiles of the different emission lines, using a Levenberg-Marquardt algorithm \citep{Levenberg1944, Marquardt1963} and the \textsc{LAPACK} Fortran libraries \citep{Anderson1999}. Extinction correction of the emission line fluxes was performed using the Balmer flux ratio $F_{H\alpha} / F_{H\beta} = 2.86$ for case B recombination \citep{Brocklehurst1971}, using the \citet{cardelli1989} extinction law.

In this first paper, we used the H$\alpha$ emission as a tracer for star formation. The detailed analysis of the ionized gas properties using other emission lines will be presented in a future paper  (Robleto-Orús et al. in prep). We applied a threshold of signal to noise ratio (S/N > 3) to the H$\alpha$ emission line in all spaxels. 

\begin{figure*}
 \includegraphics[width=0.98\textwidth]{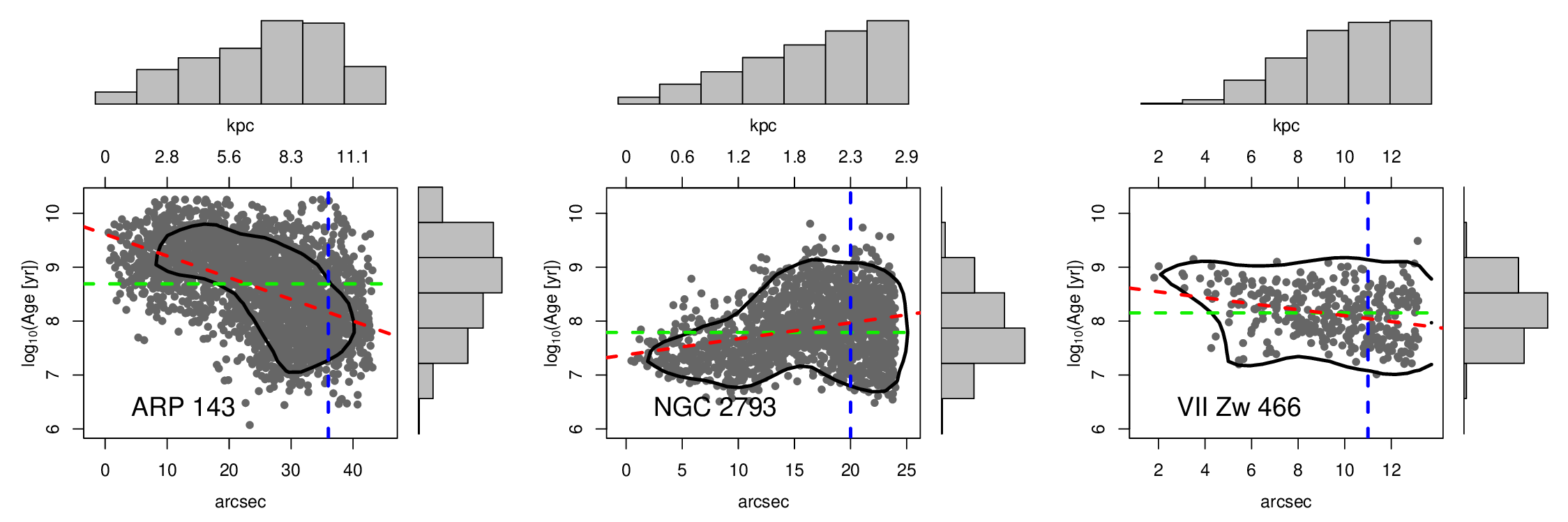}
        \caption{De-projected radius vs mean age  (light-weighted)  radial profiles for each galaxy. The radii of the spaxels were de-projected using the H$\alpha$ ellipse parameters of \citealt{romano2008}. The black contours represent the 90\% probability of finding a point. The green dashed lines represent the median age of the spaxel distributions. The red dashed lines represents the least square fit to the points. Blue vertical lines mark the semi-major axes distance (ring position). Histograms of the radial and age distribution of the spaxels are shown along box borders. Only spaxels within 1.2 times the semi-major axis distances were considered.}
 \label{fig:age_dispprof}
\end{figure*}

\section{Results}
\label{sec:res}
Figure~\ref{fig:age_map} (left panels) show synthetic RGB images of the three galaxies, produced by combining the extracted fluxes of the Johnson-Cousins $BVR$ bands from the data cubes. The H$\alpha$ luminosity is a good tracer for the ring position in this kind of galaxy, as shown by \citet{romano2008}. We plotted the H$\alpha$ emission contours in all maps for brightness at 0.1 $L_\odot$ kpc$^{-2}$ obtained from spaxels with  S/N > 3 for the line flux.

Light-weighted age $\langle \log{t_\star} \rangle_L$ and mass-weighted age $\langle \log{t_\star} \rangle_M$ maps of the three galaxies are shown in the centre and right columns of Figure \ref{fig:age_map}, respectively. For all galaxies, the location of the youngest populations in the light-weighted maps is consistent with the H$\alpha$ contours, serving as a sanity check of the procedure followed in this work. On the other hand, the mass-weighted ages are almost uniform across the disk, with these ages systematically larger than that the light-weighted ages. As results, and as it was mentioned in Section \ref{sec:SPS}, the light-weighted maps show younger mean ages than mass-weighted maps.

The difference between the light-weighted and mass-weighted ages is easily discernible in Figure \ref{fig:age_boxplot}, which shows  boxplots of the distribution of the light-weighted and mass-weighted ages for the spaxels of each galaxy. In the three cases, the median of the total distribution of light-weighted  ages ($\sim10^8$~yr) is almost two orders of magnitude lower than the mass-weighted age ($\sim10^{10}$~yr). 

In the context of the collisional scenario of ring formation, the light-weighted ages correspond to that of the populations that formed after the collision, whereas the mass-weighted ages correspond to that of the pre-collisional disk populations. In the rest of this section, we carry out a detailed analysis of the relative distribution of these two populations as a function of the distance from the ring centre.

\subsection{Age Profiles} 
In order to produce age profiles we need to consider projection effects in the plane of the sky. \citet{romano2008} accounted for these by fitting ellipses in both $B$-band (Johnson-Cousins) and H$\alpha$ images. The first one fits better the stellar disk while the second one fits better the ring position. The parameters of these ellipses (centres, semi-major axes, ellipticities and position angles) are listed in Table 4 by \citet{romano2008}. In the present work we use their H$\alpha$ image parameters to de-project the radial distances of the spaxels respect to the ring, since their images have better spatial resolution than our IFS data.

In the density wave scenario of ring formation \citep{lynds1976}, the wave expands around the impact point resulting in an expanding star formation wave. The star formation is quenched in the wake of the expanding wave due to the consumption of gas and the negative feedback from dying stars resulting in the youngest stars located at the ring position and successively older stellar populations towards the inner regions. In the more recent ring simulations by \citet{renaud2018} some of the formed stars are dragged by the expanding wave, resulting in the appearance of post-collisional stars of all ages, not just the youngest, at the ring position. In order to test these models, we need to ideally  construct age profiles radially symmetric around the impact point. However, since locating the impact point in ring galaxies is a non-trivial exercise, in this analysis we consider the centre of the ellipses as the impact point. The possibility of off-centre collisions and the nature of the ring they produce will be discussed in section \ref{sec:con}.

We used the \citet{romano2008} H$\alpha$ ellipse parameters to calculate the de-projected radial distances of the spaxels. Figure~\ref{fig:age_dispprof} shows the mean light-weighted age vs. de-projected radial distance profiles for each spaxel in the 3 galaxies within the radial distance of 1.2 times the semi-major axis of the ring. The position of the ring from \citet{romano2008} is shown by the vertical blue dashed line. For each galaxy, the plot shows the median age of all spaxels (in green) and a least squares fit (in red). Stellar populations cover a large range of ages at each radial distance. Under the collision interpretation for ring galaxies, the upper boundary correspond to the pre-collisional old disk stars and the lower boundary denote the age of the last episode of star-formation at that radius. To probe if this is the case, we need to analyze the radial gradient of the youngest populations (the ones assumed to be created during the collision, which is presented in Section \ref{sec:youngest_pop_prof}).

\begin{figure*}
 \includegraphics[width=\textwidth]{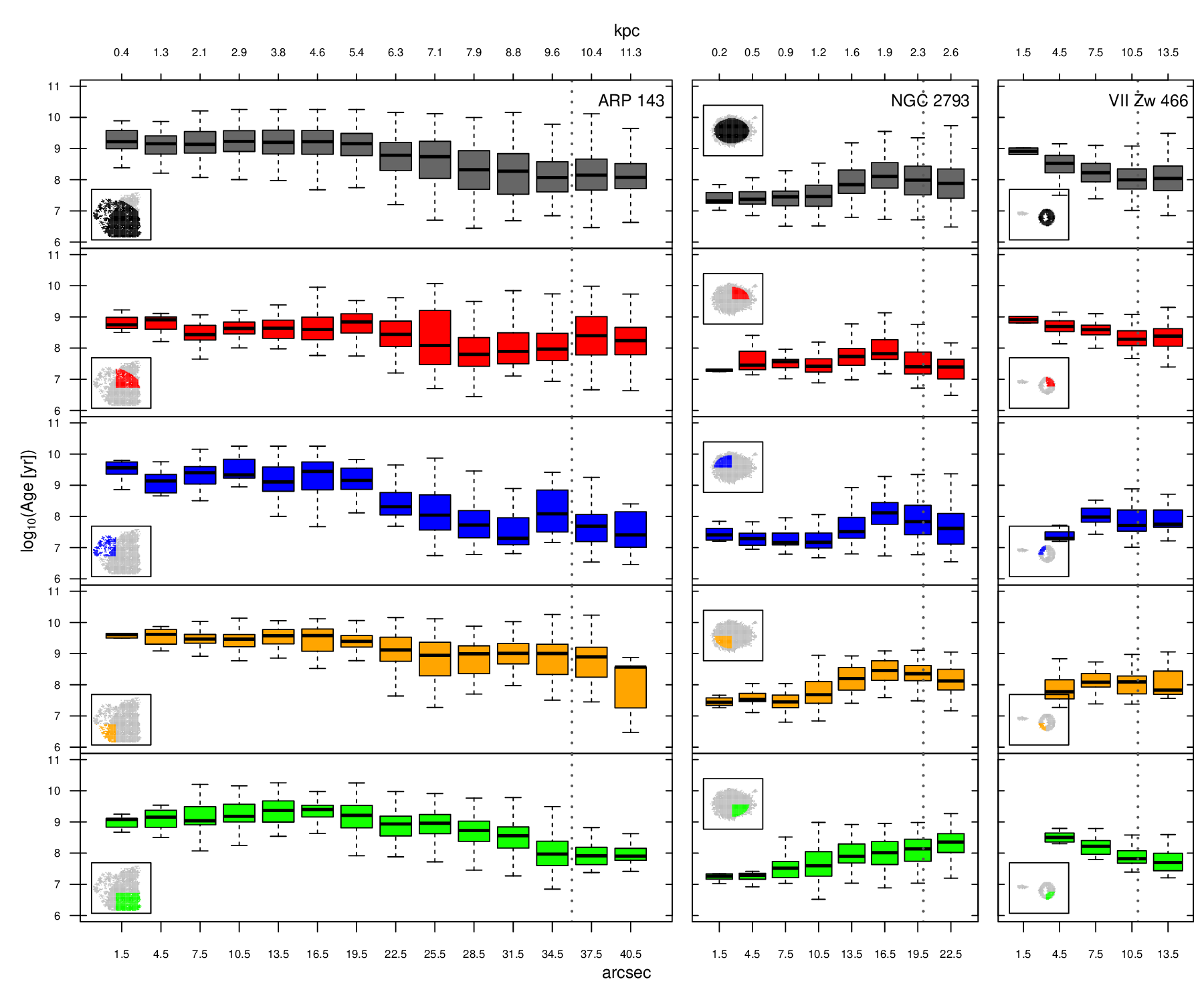}
        \caption{Box plot profiles for the light-weighted age distribution for the three galaxies. The inset maps highlight the spaxels used to produce the profiles. The top panel includes the general profile (i.e., the four quadrants of the disk together) while each of  the other panels correspond to each of the quadrants. The galactocentric radii are measured respect to the centre of the \citealt{romano2008} H$\alpha$ ellipses and the distances are de-projected from the plane of the sky. The bins have a width of $3.0$ arcsec (i.e. twice FWHM of the seeing). Box plots are drawn only if the number of spaxels in the bin is higher than five. The maximum bin reaches a distance of 1.2 times the major semi-axis radius, represented with the vertical dotted line. Boxplots are defined as in Figure \ref{fig:age_boxplot}.}
 \label{fig:age_boxprof}
\end{figure*}

In order to illustrate better the radial and azimuthal distributions of median ages of the stellar populations from the impact point up to the ring, we use box plots in Figure~\ref{fig:age_boxprof}, selecting the spaxels in bins of 3.0'' (twice the FWHM of the seeing). We restricted the analysis of the azimuthal distributions to four quadrants. The quadrants follow the Cartesian quadrants with the centres fixed to the centre of the \citet{romano2008} H$\alpha$ ellipses, which is the assumed impact point. The top-most panel for each galaxy shows the profile for the whole galaxy, which is henceforth referred to as the ``general profile'', with the bottom four panels showing it for the quadrant indicated in the bottom-left inset in each panel.

\subsection{Profiles of the youngest populations}
\label{sec:youngest_pop_prof}
Figure~\ref{fig:age_boxprof} shows that even for $\langle \log{t_\star} \rangle_L$---which should be biased towards young ages---many spaxels close to the centre of the ring, the assumed impact point,  show median ages well above $10^8$ yr, the ring lifetime expected from the models of Cartwheel-like ring galaxies \citep{renaud2018}. This is most likely due to a mix of the old populations present in the galaxy before the collision with the new populations created by the passage of the ring density wave. \citet{romano2008} arrived to similar conclusions in the case of the colour gradients. 
As the density wave generated after the ring-making impact expands, it forms stars at the position of the wave. The star formation is eventually quenched in the wake of the wave due to negative feedback from the exploding supernovae, as well as due to gas consumption \citep{korchagin2001}. Under such a scenario, the age of the youngest population at any given radius could be approximated as the epoch when the ring passed that radius, with the youngest age at the impact point corresponding to the epoch of impact. Such a scenario predicts a negative radial age gradient of the youngest populations.

\begin{figure*}
  \includegraphics[width=0.33\textwidth]{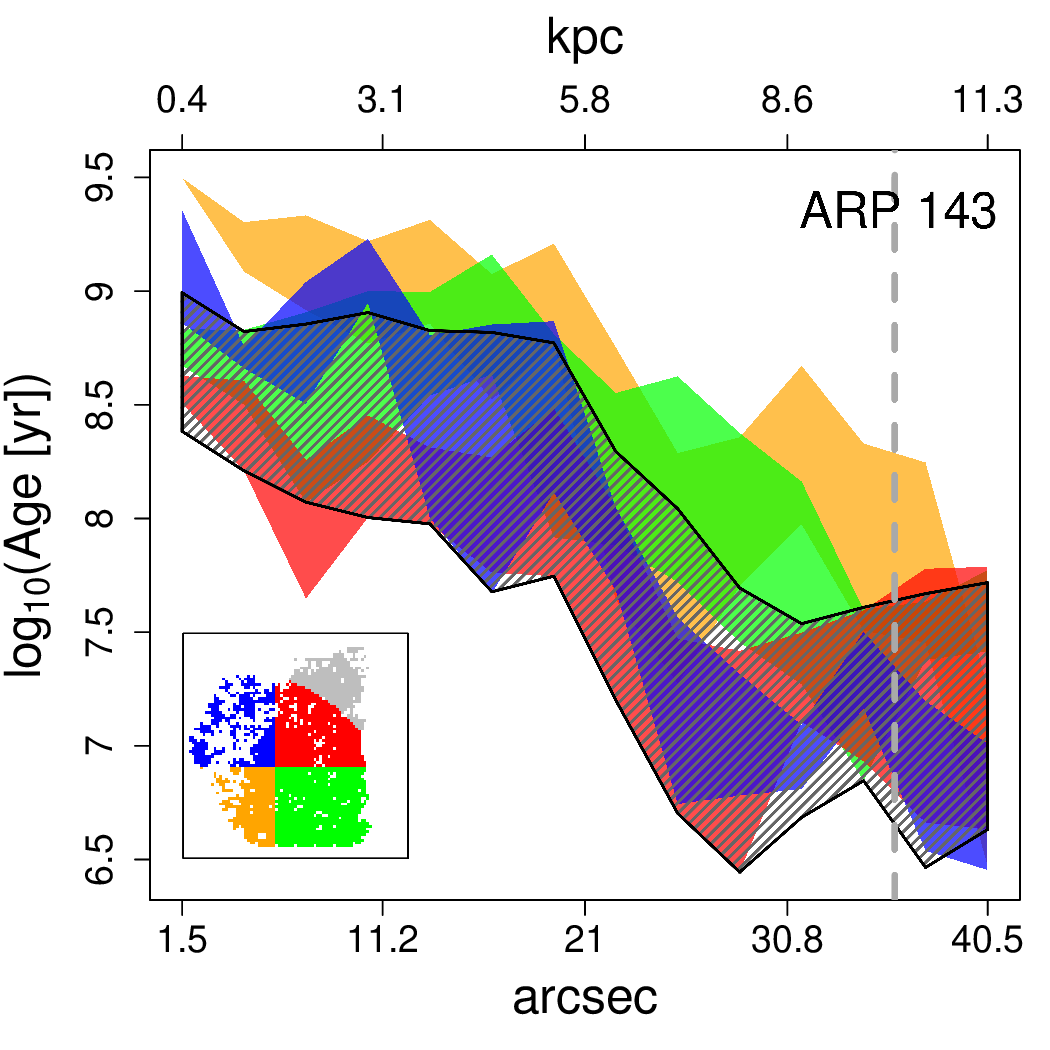}
  \includegraphics[width=0.33\textwidth]{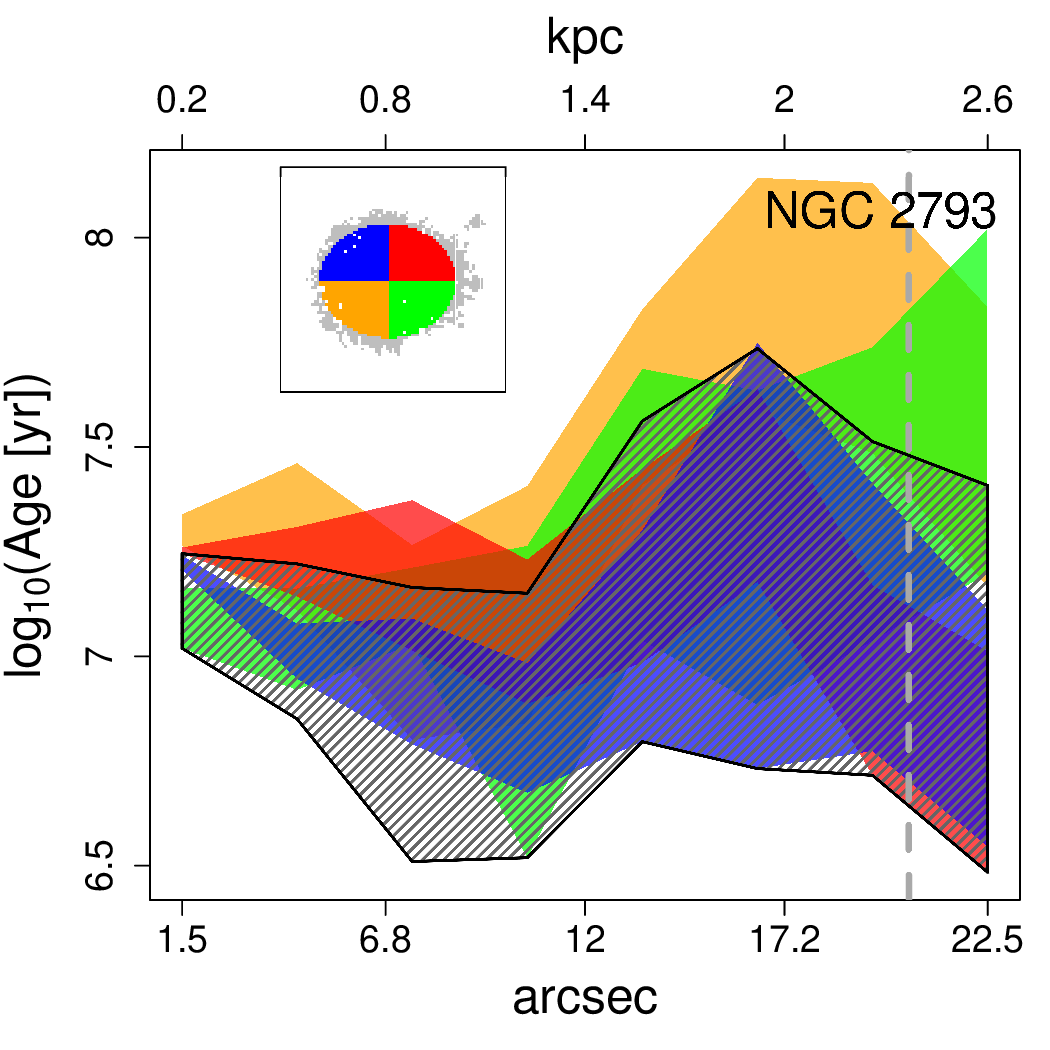}
  \includegraphics[width=0.33\textwidth]{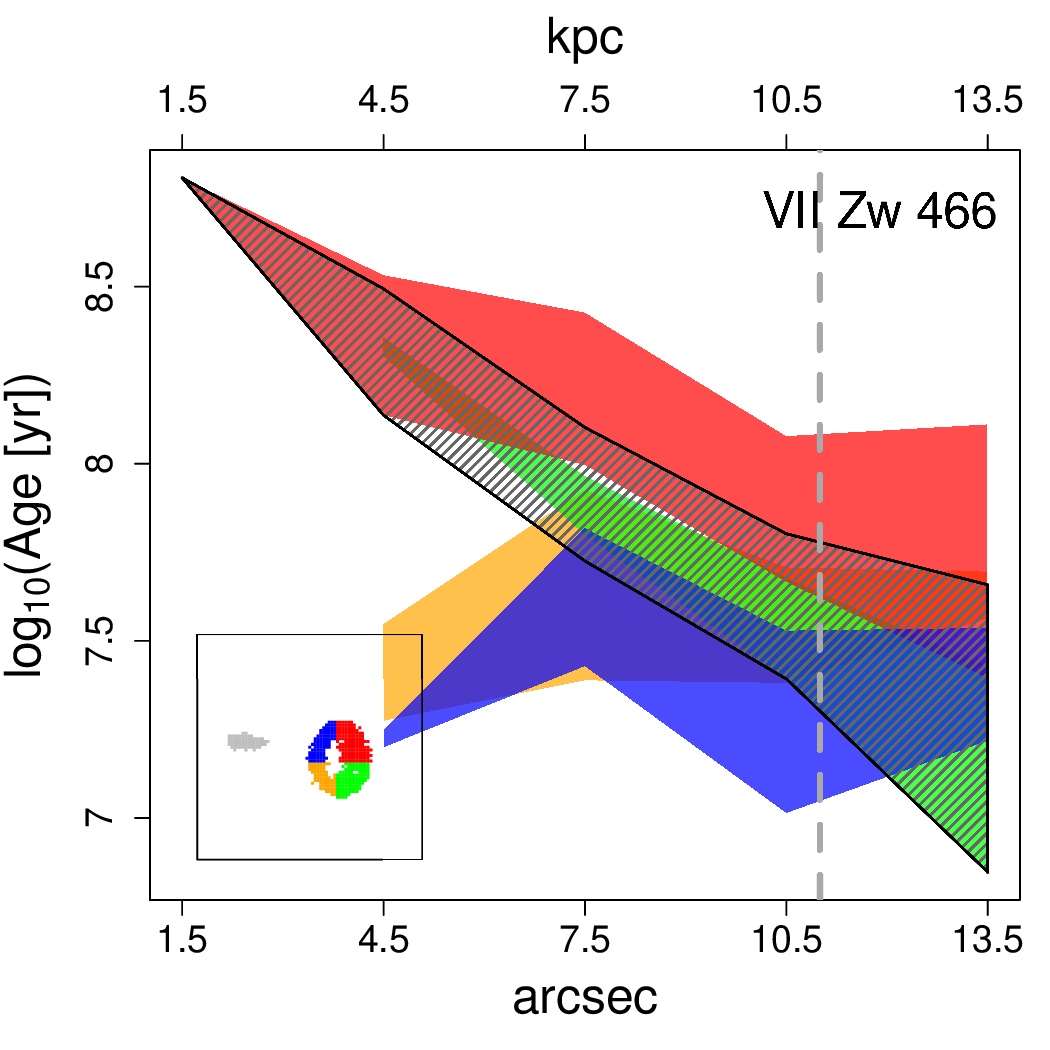}
        \caption{Profiles for the radial distribution of the youngest spaxels, i.e., any spaxel  between the lower whisker and the quartile Q1 of the box plots of figure \ref{fig:age_boxprof}. Colours indicate the profile in the direction of the corresponding quadrant indicated in the inset of each panel. The general profile is marked with hatched black. Grey dashed lines indicate the semi-major axis radius of the rings. We include the (de-projected) physical radius at the upper axis and the observed angular radius at the lower axis. For VII~Zw~466, the general profile only includes the western quadrants.}
\label{fig:age_minboxprof}
\end{figure*}

We take advantage of the boxplot profiles to isolate the spaxels dominated by the youngest populations at each radius. For this we considered only those spaxels with ages contained within the lower whisker and the first quartile (Q1) in Figure~\ref{fig:age_boxprof}. By convention, the lower whisker of a box plot begins at Q1 (top) and ends at the lowest observed data point of the distribution that is still 1.5 times the Inter-Quartile Range (IQR $= Q3 - Q1$, the range within the first and third quartiles [Q1,Q3]) from Q1 (bottom). All data points outside the whiskers boundary are considered as outliers. This is a pure statistical definition that accounts only for the distribution in each bin of the boxplots in Figure \ref{fig:age_boxprof}. The youngest populations defined in this way in each quadrant of the three sample galaxies are shown in Figure~\ref{fig:age_minboxprof}. A comparison of the ages of the youngest population at the normalized radius in the three galaxies is carried out in Figure~\ref{fig:com_age_min}.

Once we have the profiles, we define the gradient of the mean age of the youngest populations  (as defined above) as:

\begin{equation}
        \label{eq:gradient}
        {\nabla t_\star}_y= \frac{\langle \log{t_\star} \rangle_y^{R=R_{ring}}-\langle \log{t_\star} \rangle_y^{R=0.5R_{ring}}}{0.5 \ R_{ring}}
\end{equation}

\noindent
where $\langle \log{t_\star} \rangle_y^{R=R_{ring}}$ and $\langle \log{t_\star} \rangle_y^{R=0.5R_{ring}}$ are the mean ages of the ``youngest'' populations, for the bin at the ring location and for half the ring radius, respectively. Both values were obtained by interpolating the ages of the spaxels contained by the two bins closest to such locations. We computed the mean gradient directly instead of deriving it from linear regression because we are analysing not a dispersed set of points but the extreme values of the lower whisker in boxplots. Calculating the difference between two well-separated points provides a more robust estimate, as intermediate points could be influenced by internal trends within the profiles.

The choice of determining the gradient between $R=0.5R_{ring}$ and $R_{ring}$ rather than the entire extent of the ring is based on the following three reasons: 1) the difficulty in isolating the youngest population close to the central regions as they are affected by older populations by a greater extent as compared to outer regions, 2) the central regions have lesser number of spaxels for a statistically significant analysis, and 3) the assumption of the ellipse centre as the impact point makes a lesser effect at outer radii, especially in small angular sized galaxies such as VII~Zw~466. The quantity ${\nabla t_\star}_y$ has units of $\log_{10}(Age[yr])$, and coincides with the gradient if we use the normalized radius at $R_{ring}$. 

\begin{table*}
\caption{\label{tab:gradients} ${\nabla t_\star}_y$ taken from Figures \ref{fig:age_minboxprof}; for the general gradient and for each quadrant in the anti-clockwise sense, where colour description coincides from such figures. All gradient units are in $\log_{10}$(yr)}
    \centering
    \begin{tabular}{lcccccc}
        \hline
        Galaxy & $R_{Ring}$ & ${\nabla t_\star}_y$ & ${\nabla t_\star}_{y1}$ & ${\nabla t_\star}_{y2}$ & ${\nabla t_\star}_{y3}$ &  ${\nabla t_\star}_{y4}$\\
        & (kpc) & (black) & (red) & (blue) & (orange) & (green) \\
        \hline
        Arp~143  & 10.0  & $-2.2\pm1.0$ & $-1.6\pm0.8$ & $-2.6\pm0.7$ & $-2.0\pm0.7$ & $-2.5\pm0.6$ \\
        NGC~2793 & 2.3  & $0.5\pm0.7$ & $-0.3\pm0.4$ & $0.4\pm0.5$  & $1.3\pm0.6$  & $1.0\pm0.7$ \\
        VII~Zw~466 & 11.0 & $-1.3\pm0.4$ & $-0.9\pm0.4$ & $-0.1\pm0.3$ & $0.1\pm0.3$  & $-1.4\pm0.2$          \\
        \hline 
    \end{tabular}    
\end{table*}

Table \ref{tab:gradients} lists the values of ${\nabla t_\star}_y$ for the whole galaxy profiles and for the four quadrants for each galaxy. In the case of VII~Zw~466, as mentioned previously, only western quadrants were used to compute ${\nabla t_\star}_y$, despite the values of each quadrant were indicated in the table. In general, Arp~143 and VII~Zw~466 shows the most significant gradients, both with ${\nabla t_\star}_y\leq-1.0$, while NGC~2793 is consistent with a flatter gradient.

We use the plots in Figures~\ref{fig:age_boxplot} to \ref{fig:com_age_min} to discuss the three galaxies under the collisional scenario of ring formation following the classical models of \citet{theys1976} and the more recent simulation of \citet{renaud2018} in the next subsection.

\subsection{Ages of post-collisional stellar populations in the sample galaxies}

\subsubsection{Arp~143} 
Arp~143 exhibits a ring formed by knots of star formation and disrupted material (see Figure~\ref{fig:age_map}). Notably, the B-band emission previously reported by \citet{romano2008} as tracer for stars in the disk, reaches larger galactocentric distances than the H$\alpha$ emission. This B-band emission extends beyond our field of view. Nevertheless, the star formation ring traced by the H$\alpha$ emission is mostly contained within our field of view except for parts of the north-east quadrant, marked by the green ellipse in Figure~\ref{fig:age_map}. The companion is located to the north-west (outside of the observed field) and the original bulge of Arp~143 is located in the inner region, although presenting a slight offset toward the companion with respect to the centre of the ring.

Consistent with \citet{romano2008} and \citet{higdon1997}, we also detected H$\alpha$ emission from the nucleus, pointing to the existence of star formation in that region. \citet{higdon1997} suggested that the collision could have produced a long H I tidal tail that funnels gas toward the centre of Arp~143, which could be  fuelling the star formation. The fact that the apparent bulge has an offset respect to the centre of the ring is consistent with an off-centred collision scenario, resulting in important asymmetries in the age maps (see Figure \ref{fig:age_map}).

The boxplot profiles present negative gradients for all quadrants, being the north-west quadrant (red) the one with the flattest gradient and the north-east quadrant (blue) the steepest.
The lower boundary of points for Arp~143 (Figure~\ref{fig:age_dispprof}) present a much steeper negative age gradient as compared to the median ages, with the populations being the youngest at the position of the ring, as expected for ring galaxies. A systematically ageing of the population towards the centre of the ring is easily noticeable in radial profiles for all the four quadrants. The youngest ages for the four quadrants at the same radius span almost an order of magnitude spread with the ages in the north-western quadrant (red) being systematically younger. The systematic difference of ages in the four quadrants at a given radius suggests different amounts of contamination from pre-collisional disk populations in different quadrants. We take the profile of the quadrant with the youngest ages to determine the minimum age elapsed since the collision. This age corresponds to 300~Myr at the ring centre, reaching 100~Myr and 50~Myr at 25\% and 50\% ring radius. 

\begin{figure}
  \includegraphics[width=0.99\columnwidth]{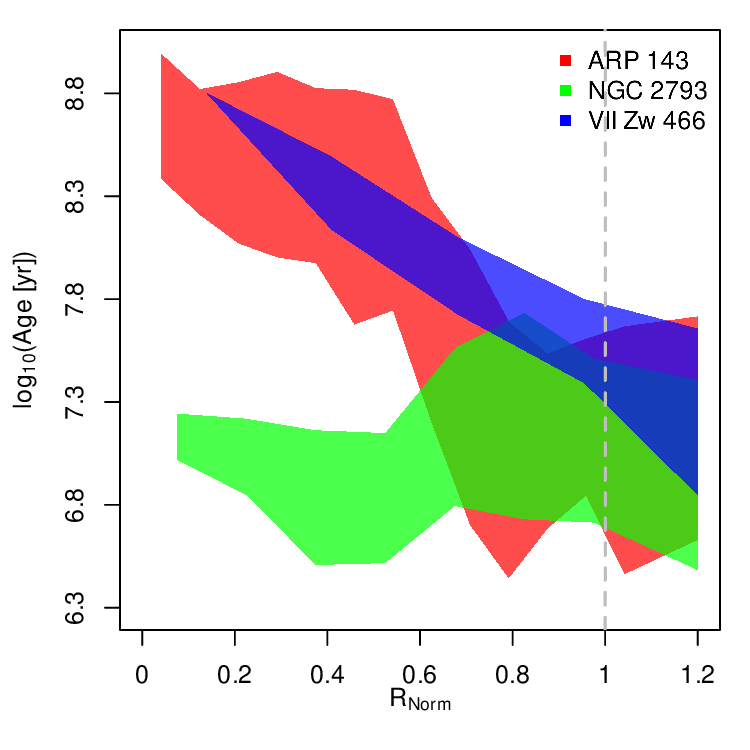}
        \caption{Comparison of youngest population age profiles for the three galaxies, with their radii normalized at the ring position ($\textrm{R}_{\textrm{Norm}}$).}
\label{fig:com_age_min}
\end{figure}

\subsubsection{NGC~2793}

With a mass of $4\times10^{9}$~M$_\odot$ \citep{romano2008}, NGC~2793 is the lowest mass ring galaxy candidate of our sample. The physical size of the ring in NGC~2793 is only 2.1~kpc, in spite of its relatively large angular size of 22.5'' (the second largest of our sample). The physical size of the galaxy (4.2~kpc) is also small and is typical of a dwarf galaxy. 

The inner regions of NGC~2793 are characterized by young and intermediate age populations (Figure~\ref{fig:age_map}). This is consistent with \citet{romano2008}, who found NGC~2793 to be the galaxy with highest B-band intensity of their sample in the inner regions ($>$0.2 mag at the centre). The supposed ring-structure is clearly traced by the H$\alpha$ emission except for the south-west quadrant, where it is almost absent. The age profile of NGC~2793 (Figure~\ref{fig:age_dispprof}, centre) presents a positive gradient inconsistent with the collision hypothesis. 

The age profile in NGC~2793 is dominated by spaxels with ages of $10^{7-7.3}$ yr at the first bin (1.5'') in Figure \ref{fig:age_boxprof} (top panel) , but the dispersion increases and older ages dominate at larger radii. The IQR goes from 0.4 dex at the first bin to 0.9 dex at the 22.5~arcsec bin. 
There is also a little increment in the median, from $10^{7.3}$ to $10^{7.8}$yr, consistent with the positive age gradient in Figure \ref{fig:age_dispprof}. The boxplot profile shows that it is almost flat in the inner regions up to the 13.5 arcsec bin, when it ascends up to a maximum at the 16.5~arcsec bin (where the IQR is 0.8 dex). This maximum is  therefore the responsible for the positive gradient in Figure~\ref{fig:age_dispprof}.

The maximum at the 16.5~arcsec bin is visible in the general profile (dark grey in Figure~\ref{fig:age_boxprof}), in the north-west (red), in the north-east (blue), and in the south-east (orange) quadrants. The south-west (green) quadrant is the only one that does not show this feature in its profile, but anyway the stellar age gradient is clearly positive there too. Remarkably, the youngest age profiles in Figure \ref{fig:age_minboxprof} (centre), follow a similar trend to the boxplot profiles in Figure \ref{fig:age_boxprof}.

It is not obvious from the RGB images (left panels of Figure~\ref{fig:age_map}) nor from the age maps (right panels of Figure~\ref{fig:age_map}) what is causing the maximum. The ring position, as determined by \citet{romano2008} is located at 20~arcsec, so this maximum does not overlap with the ring. In Figure \ref{fig:age_map} we can see what appears to be the bulge close to the east part of the ring, which could explain that this increment in age is the highest for the south-east quadrant (orange in Figure \ref{fig:age_minboxprof}.), but this should have no effect in the western quadrants.

We also want to note that in the south-west (green) quadrant (the only one without the maximum) the H$\alpha$ emission is scarcer than in the other quadrants. Considering this, we propose a possible explanation for this stellar age maximum: The whole galaxy formerly had a positive stellar age gradient throughout the outer regions, alike to the south-west quadrant. Afterwards, when the current star formation burst appeared in the outskirts, it caused the median age values to go down in those regions, consequently producing a local descend in the profile. This scenario could effectively explain the maximum observed in the two eastern quadrants and in the north-west quadrant (i.e., where the H$\alpha$ emission is prominent). As for the south-west quadrant---devoid of  prominent H$\alpha$ emitting regions---the recent star formation burst probably did not have a significant effect there, preserving the original positive gradient all the way outwards, as observed. Although most low-redshift dwarf galaxies show flat to slightly negative stellar age gradients---both in the inner and outer regions---cases with positive gradients also exist (e.g., see Figure 8 by \citealt{cano-diaz2025}).

Regarding the increment in the IQR towards larger radii, it could be explained by the increment in the number of spaxels used in each bin.

Observed star formation characteristics, namely the presence of star formation inside the ring and the positive age gradients, rule out this galaxy as a genuine collisional ring galaxy, in spite of a ring present in the H$\alpha$ image. 

\subsubsection{VII~Zw~466}
\begin{figure}
 \includegraphics[width=0.97\columnwidth]{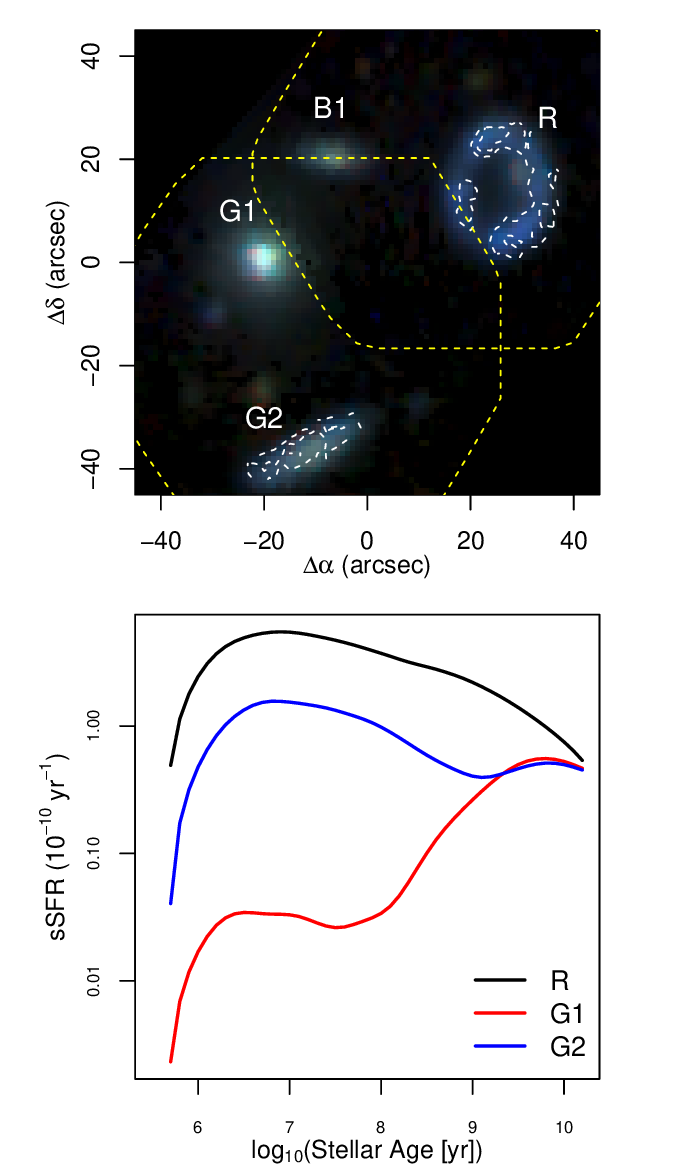}
        \caption{Top: RGB image of the VII~Zw~466 group, based on the IFS data, including the two overlapped fields of view (yellow dashes lines). The IDs are consistent with \citet{romano2008}. The NED Identifiers for each galaxy are shown in Table \ref{tab:viizw466}. Bottom: Specific star formation histories (sSFH), calculated as the median of the whole spaxels by each galaxy. The horizontal axis shows the stellar age at the moment that the observed light was emitted. Both axes are in logarithmic scale.}
 \label{fig:ssfh_viizw466}
\end{figure}

VII~Zw~466 is part of a system formed by (apparently) four galaxies and shows an evident ring morphology (Figure~\ref{fig:age_map}, bottom panels). 
In order to unveil which of the four galaxies was the one that collided with VII~Zw~466, we captured the whole system with two fields of the instrument, with part of them overlapping (see the top panel of Figure~\ref{fig:ssfh_viizw466}). The group is catalogued by the Uppsala General Catalogue of Galaxies \citep{nilson1973} as UGC~07683. 

Despite VII~Zw~466 showing the smallest angular size of the three galaxies, it presents the larger H$\alpha$ luminosity.  Up to 92\% of the H$\alpha$ emission originates in the western half of the ring. A small region with redder colours and mass-weighted ages $\leq 10^{10}$ yr is present in the inner border of the west side of the ring, which could be associated with the bulge of the original disk galaxy. 
 
This galaxy has the most complete ring in the sample, with a semi-major axis distance of 9.3~arcsec. With the $S/N>10$ criterion for the continuum, many spaxels in the inner regions get excluded. Indeed, in \citet{theys1976}, VII~Zw~466 was classified as a ``RE'' galaxy, i.e., elliptical ring with empty interior. VII~Zw~466 shows a clear negative age gradient, as expected from a collision scenario.  

The interior of the ring appears empty, with only the red quadrant containing a bin at 1.5~arcsec. In its general profile (Figure \ref{fig:age_boxprof}) VII~Zw~466 exhibits a negative gradient. This trend is also observed in the western quadrants (red and green) but not in the eastern ones (blue and orange), where the gradient appears to be flat. The absence of a gradient in the eastern quadrants may result from the limited number of spaxels with S/N$\geq$ 10 available to resolve the profile in that direction, in contrast to the western quadrants. Consequently, these regions are not representative of the ring and would not be suitable for gradient determination. Therefore, we exclude the eastern quadrants from further computations.

VII~Zw~466 has a steep youngest-population profile ${\nabla t_\star}_y$ (see Table \ref{tab:gradients}). As explained before, and despite all quadrants were taking into account in Figure~\ref{fig:age_boxprof} (right-panel) and in Table \ref{tab:gradients}, only western quadrants were considered for the general profile. Indeed, the profiles of the eastern quadrants (Figure~\ref{fig:age_minboxprof}, blue and orange) appear to be flatter.  Due to the projection in the line of sight, we are probably observing only the outer side of the ring of these quadrants, which could explain why we observe younger ages. As a result, the general profile appears to be similar to the western profiles, which are more consistent with that produced after the collision.

The western quadrants (red and green) present more dramatic gradients, with values of ${\nabla t_\star}_y$ $\sim-1.0$ and minimum age of $\sim 10^{7.5}$ yr. The first quadrant (the red one) is hosting what appears to be the original bulge (Figure~\ref{fig:age_map}). This could introduce older populations from the bulge. Despite that, the fourth quadrant (green one), where we do not expect bulge contamination, has even a larger negative value of ${\nabla t_\star}_y$.

The youngest populations close to the ring centre in the western half suggests the post-collisional star formation happened around 100--300~Myr ago. In comparison, the star formation in the eastern half stopped as recently as $\sim$20~Myr ago.

\subsection{Profile Comparison}
Figure~\ref{fig:com_age_min} shows a comparison of the general profiles of the three galaxies normalizing the radii by the semi-major axis of their rings ($\mathrm{R_{Norm}}$). The figure shows that Arp~143 and VII~Zw~466 profiles have similar steep value of ${\nabla t_\star}_y$ of -2.2 and -1.3 respectively), both doubling the error range. It is important to note that the youngest population at low $\textrm{R}_{\textrm{Norm}}$ are old in comparison with the estimated lifetime for the ring from simulations ($\sim 10^{7}$ yr). For both Arp~143 and VII~Zw~466 it is obvious that the youngest stars of the whole galaxy are located in the rings. This is consistent with the collision hypothesis, and, indeed, the internal regions seem to have halted the star formation after the passage of the density wave of the ring. 

Meanwhile, the general profile for NGC~2793 (Figure~\ref{fig:age_minboxprof}, centre) is almost flat (${\nabla t_\star}_y$ = 0.5$\pm$0.7). Comparing with the other two galaxies (Figure~\ref{fig:com_age_min}), this flatness is evident. Such flat profile is not consistent with the collision hypothesis. Therefore, NGC~2973 is probably not a collisional ring galaxy.

\subsection{Expansion velocity of the wave}
\label{sec:expansion}

Obtaining expansion velocities of density waves in ring galaxies is a challenge  \citep{vorobyov2003}. Previous attempts to measure expansion velocities use gas velocities either from CO or HI lines from the ring \citep[e.g.,][]{appleton1996,higdon1997}. The spatially-resolved ages provide an alternative method to infer the expansion velocity.

The lower limit for the expansion velocity of the wave can be computed by assuming that the age of the youngest population at the impact point (here assumed as the centre of the ellipses, see Section\ref{sec:ColGeom}) corresponds to the epoch of the collision and that the wave has been expanding uniformly since then.
With an age of the youngest population at the impact point of 300~Myr and ring radius of 10~kpc for Arp~143, and 100~Myr and 11~kpc for VII~Zw~466, we estimated expansion velocities of  33$\pm$15~km~s$^{-1}$ an  108$\pm$30~km~s$^{-1}$, respectively.

For both galaxies, we find that our measured velocities differ from those reported in the literature. In the case of Arp~143, our value is significantly lower than the expansion velocity of 118$\pm30$~km~s$^{-1}$ of the CO gas in the ring reported by \citet{higdon1997}.

In the case of VII~Zw~466, our velocity estimation of 108$\pm$30~km~s$^{-1}$ is about three times higher than the one by \citet{appleton1996}. They reported an expansion velocity of 32~km~s$^{-1}$ for VII~Zw~466 using the HI gas in the ring. In this case, the authors did not provide an uncertainty estimation.

Stellar velocities estimated in this work trace the time averaged velocity of the propagating star formation wave, whereas the velocities of the gas in the ring trace the instantaneous velocity of the expanding density wave. For a uniformly expanding density wave, the two velocities are expected to coincide. The differences suggest non-uniform expansion of the density waves. It should also be considered that the radio beams used for measurement of gas velocities are large, so as to be affected by effects of stellar feedback. Higher spatial resolution observations in both these tracers are required to carry out an elaborative discussion on the nature of the expansion of the density wave.

\begin{table}
        \centering
            \caption{Parameters derived from the analyses performed in this study.}
        \label{tab:result}
        \begin{tabular}{lcccc} 
        \hline
        Galaxy  &   ${\nabla t_\star}_y$     &  Collision Age  & \multicolumn{2}{c}{Expansion Velocity} \\
                &   $log_{10}(yr)$           & (Myr) ago      & (km s$^{-1}$) & (kpc Gyr$^{-1}$)\\
        \hline
        Arp~143    &  -2.2$\pm$1.0       &  300  & 33$\pm$10 & 33$\pm$10 \\
        NGC~2793   &   $\sim$0.0$\pm0.5$ & -   & -  & -  \\
        VII~Zw~466 &  -1.3$\pm$0.4       & 100 & 108$\pm$28 & 110$\pm$29 \\
        \hline
        \end{tabular}
\end{table}

\subsection{The companions of VII~Zw~466}
\label{sec:viizw466}

As mentioned, VII~Zw~466 is part of a group of apparently four galaxies known as UGC~07683 \citep{nilson1973}. While the individual galaxies are registered in the NASA/IPAC Extragalactic Database with proper names, we followed the \citet{appleton1996} nomenclature to label the group members G1 and G2, and the background galaxy B1 while we assigned the ID R to VII~Zw~466. The NED identifiers, ID and properties of the field galaxies are shown in Table~\ref{tab:viizw466}.

We observed the whole system using two pointings of the telescope (with their fields of view overlapped) to determine which of the galaxies is the one that collided with VII~Zw~466. The complete group of VII~Zw~466 is shown at the top panel of Figure~\ref{fig:ssfh_viizw466}. This RGB reconstruction was obtained similarly to the RGB images in Figure~\ref{fig:age_map}. 

We determined the optical continuum and emission radial velocities of the system ($v_c$ and $v_e$, respectively by integrating the central spaxels (SN $\geq$ 30) of each galaxy in a radius of 3''. The measurements were obtained using the \textsf{xcsao} and \textsf{emsao} tasks from the \textsf{rvsao} package \citep{mink1998} within \textsf{IRAF} \citep{tody1993}. These tasks perform cross-correlation between our spectra and a library of extragalactic objects to account for the continuum. Whenever possible, $v_e$ was directly estimated by fitting the emission lines. The results are presented in Table~\ref{tab:viizw466}.

The optical velocity we derived for VII Zw 466 is consistent with the HI velocity reported by \citet{appleton1996}. We also measured optical velocities for galaxies G1 and B1, which lack previously reported HI emission. Our results represent a significant improvement over the approximate optical velocities from \citet{theys1976} for these two galaxies. In agreement with their work, we find that B1 exhibits a significantly different radial velocity compared to the other members of the system. This discrepancy confirms that B1 is a background galaxy rather than a gravitationally bound component of the group.

On the other hand, for the suspected companion G2, our velocity is 14224~km s$^{-1}$, which is clearly outside the \textcolor{blue}{14334--14516}~km s$^{-1}$ velocity range  reported by \citet{appleton1996}. Interestingly, G2 reappears in their HI channel map (their Figure 8) in just one channel (14221~km s$^{-1}$) consistent with our optical velocity. This points to the possibility that the HI structure associated to G2 by \citet{appleton1996} was part of the plume connecting G2 with VII Zw 466, and that the real velocity of G2 was the one measured by the H$\alpha$ line. This would make G2 differing by as much as 350~km s$^{-1}$ in velocity with respect to the ring galaxy, suggesting an impact at higher velocity than previously thought.

\citet{appleton1996} suggested G2 as the most likely intruder. However, they did not rule out the possibility of G1 playing some role. To investigate this possibility, we analyse the specific star formation histories (sSFH) of the three group members R, G1, and G2. The sSFH traces the evolution of the specific star formation rate (sSFR; star formation rate per unit stellar mass) of the galaxy over time. Following \citet{asari2007}, the sSFR is defined as:

\begin{table*}
        \caption{Galaxies in the group of VII Zw466. ID is the same as in Figure~\ref{fig:ssfh_viizw466} for consistency with \citet{romano2008}, as explained in the text. The radial velocities $v_c$ and $v_e$ are the continuum and emission lines (when possible) radial velocities, respectively, extracted from the integration of the central spaxels inside a radius of 3''. The estimation of $v_c$ and $v_e$ was obtained using the \textsf{xcsao} and \textsf{emsao} task of \textsf{rvsao} package \citep{mink1998} from IRAF software \citep{tody1993}. Mean Age is the median age for all spaxels with SN$\geq$10 in the continuum in each galaxy. 
       }
 \label{tab:viizw466}
 \begin{tabular}{lccccccc}
  \hline
         Galaxy NED Identifier & ID &  RA & Dec &  Mean Age   & $v_c$ & $v_e$ \\
                               &    &     &     &  $\log(yr)$ & km s$^{-1}$ & km s$^{-1}$ \\
  \hline
         VII~Zw~466              & R  & $12^{h}32^{m}04.4^{s}$ & $+66^{\circ}24'16''$ &  8.15$\pm$0.12 & 14573$\pm$14 & 14530$\pm$ 10  \\
         UGC 07683 NOTES02 NED02 & G1 & $12^{h}32^{m}13.2^{s}$ & $+66^{\circ}23'59''$ &  
         9.77$\pm$10 & 13958$\pm$12 & $-$   \\
         UGC 07683 NOTES02 NED03 & G2 & $12^{h}32^{m}11.7^{s}$ & $+66^{\circ}23'22''$ &  
         8.81$\pm$14 & 14224$\pm$ 2 & 14225$\pm$ 24 \\
         UGC 07683 NOTES02 NED01 & B1 & $12^{h}32^{m}10.8^{s}$ & $+66^{\circ}24'19''$ &  9.00$\pm$0.30 & 25008$\pm$10 & 24993$\pm$151   \\
  \hline
\end{tabular}
\end{table*}

\begin{equation}
        \label{eq:ssfh}
        \textrm{sSFR}(t_\star)=\frac{1}{M^c_\star}\frac{dM^c_\star(t_\star)}{dt_\star}\sim\frac{\log e}{t_\star}\frac{\mu^c_s(t_\star)}{\Delta\log t_\star}
\end{equation}

\noindent
where $t_\star$ is the age of the SSP, sSFR$(t_\star)$ is the specific star formation rate at $t_\star$, $M_\star^c (t_\star)$ is the mass of gas converted into stars at that age, and $\mu^c_s(t_\star)$ is the fraction of the stellar mass accumulated over the galaxy's history at that age.


Since the three galaxies R, G1 and G2, are at about the same redshift, they are observed at the same lookback time. This allows to directly compare their star formation histories with their stellar ages being consistent among them. In addition, the sSFR was convoluted with a Gaussian filter in $\log t_\star$ with a FWHM of 1 dex, in order to smooth the sSFH curves  accounting for the relatively high uncertainties in time of the population synthesis technique.

These curves are shown at the bottom panel of Figure~\ref{fig:ssfh_viizw466}, where the sSFR of the three galaxies is plotted against the stellar age. The plotted sSFHs were taken as the median by bin of the sSFH from all the spaxels of each galaxy. The curve for G1 (red) clearly differs from the other two galaxies. At large stellar ages ($\sim10^{10}$ yr), the three galaxies had similar sSFR. This is not surprising since they form a group, and they are expected to originate and evolve in a common environment.

At more recent stellar ages their evolutionary paths diverge. The elliptical companion galaxy G1 reaches the peak of sSFR large stellar ages, early in its formation history, with hardly any recent star formation. This star formation history is typical of early-type galaxies observed in surveys such as SDSS and CALIFA \citep[i.e.,][]{asari2007,cidfernandes2013}. 

On the other hand, both the ring galaxy (black) and the suspected companion G2 (blue) reach their highest peak of sSFR at more recent times ($< 10^{9}$ yr ). The sSFR for the ring galaxy and G2 start  deviating strongly from that for G1 at a stellar age of a few hundreds of million years, which is consistent with the collision epoch derived from radial age profiles (Figure~\ref{fig:age_boxprof}). 

The plot points to both R and G2 passing through a recent episode of high star formation that did not affect G1. The temporal synchronization of the sSFR curves of these two galaxies  reveal compelling evidence for a common recent evolution, likely triggered by the same event: the collision.

\section{Discussion}
\label{sec:con}

\subsection{On the collision geometry}
\label{sec:ColGeom}
As mentioned in section \ref{sec:method}, we assume a centred collision with the impact point close to the centre of mass, implying that the centre of the ring is also the impact point. Scenarios with the compact galaxy impacting the disk galaxy off-centre were considered in simulations by \citet{toomre1978} and were explored in more detail by \citet{appleton1987}. Their results show that in case of an off-centre collision, the shock wave is expected to produce a ring with azimuthal asymmetries in the star formation rate, H$\alpha$ luminosity, etc.. As a result, rings could have a radially asymmetrical appearance, where the collision effects will be more important in some directions than in others.

For our sample, the three galaxies show important asymmetries in the H$\alpha$ emission---and hence in star formation---distribution along the rings, evident from the contours in Figure~\ref{fig:age_map}. In the case of Arp~143 these asymmetries were studied by \cite{beirao2009} using Spitzer IR bands and GALEX UV bands. They confirm the origin of the star formation knots of the ring as being simultaneously produced by a shock wave detected with H$_2$ emission. They find their results consistent with the off-centre collision scenario by \cite{appleton1987}, with the formation of the ring resulting from a combination of an expanding wave and a disk rotation movement.

That said, we want to argue in favour of our decision to assume the centred collision scenario to measure the gradients. For Arp~143 the youngest age profiles in Figures~\ref{fig:age_minboxprof}
and \ref{fig:com_age_min} show that the oldest among the youngest spaxels are located closer to the origin, producing the negative age gradient that we report. In the case of VII~Zw~466, we see the same trend for the western quadrants (red and green in Figure~\ref{fig:age_minboxprof}) with the oldest among the youngest ages being closer to the centre of the ring and a clear negative age gradient, consistent with the centred-collision hypothesis. In the case of the eastern quadrants, the lack of spaxels at the inner regions prevent us to obtain conclusions. Since we no longer consider NGC~2793 as a collisional ring, we will not discuss its case. 

Considering other kinds of asymmetries is also relevant for our sample. \cite{toomre1978} showed that off-centre collisions produce rings with the centre separated from the original bulge of the disk galaxy. The possible original bulge of any of our two collisional ring galaxies is not exactly at the centre of its ring. Therefore it is possible that the intruder galaxy did not impact properly in the ring centre in any of the two cases. In the case of Arp~143, the bulge is slightly de-centred. This points to the possibility that even if the impact was off-centre, it was only slightly so (similar to row 5 in Figure 5 by \citealt{toomre1978}). 

In addition, we are limited by the spatial resolution of our data ($\sim~1.5"$ seeing and $1"$ spaxel size), comparable with the projected distances from the bulges of the two galaxies to the centres of their rings (a few arcseconds); this could effectively blur the effects of this asymmetry on the gradients for the central regions.

In conclusion, given the clear negative gradients of the youngest age profiles and the limitations in spatial resolution of our data, the centred collision scenario is good enough for our purpose. Further studies with higher spatial resolution and better spectral resolution (to determine kinematics) could help to localize the impact point, as suggested by \cite{appleton1987}.

\begin{figure*}
  \includegraphics[width=0.95\textwidth]{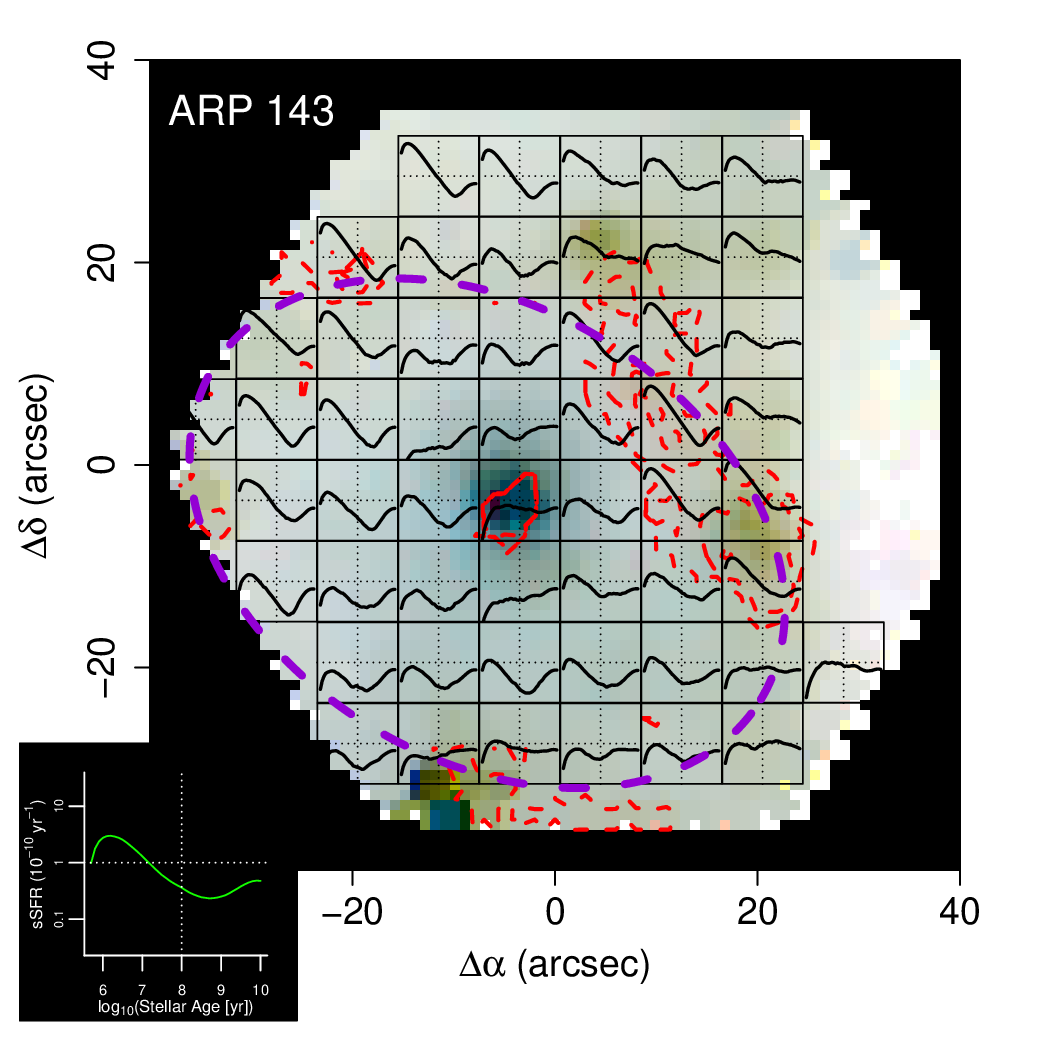}
        \caption{sSFH functions on the RGB negative map of Arp~143. Each box shows the median sSFH function produced by the contributing spaxels, drawing on them. The contribute spaxels correspond to an $8\times8$ array of spaxels, just where the box is located. S/N criterion was relaxed to incorporate all spaxels on the internal region of the ring. The scale of these plots is shown on the bottom-left corner, also showing the median sSFH of the all spaxels of the galaxy (in green). H$\alpha$ luminosity levels are showed in red and \citet{romano2008} ellipse is showed in violet. sSFH plots axes are in logarithmic scale.}
\label{fig:arp143_ssfh}
\end{figure*}

\begin{figure*}
  \includegraphics[width=0.95\textwidth]{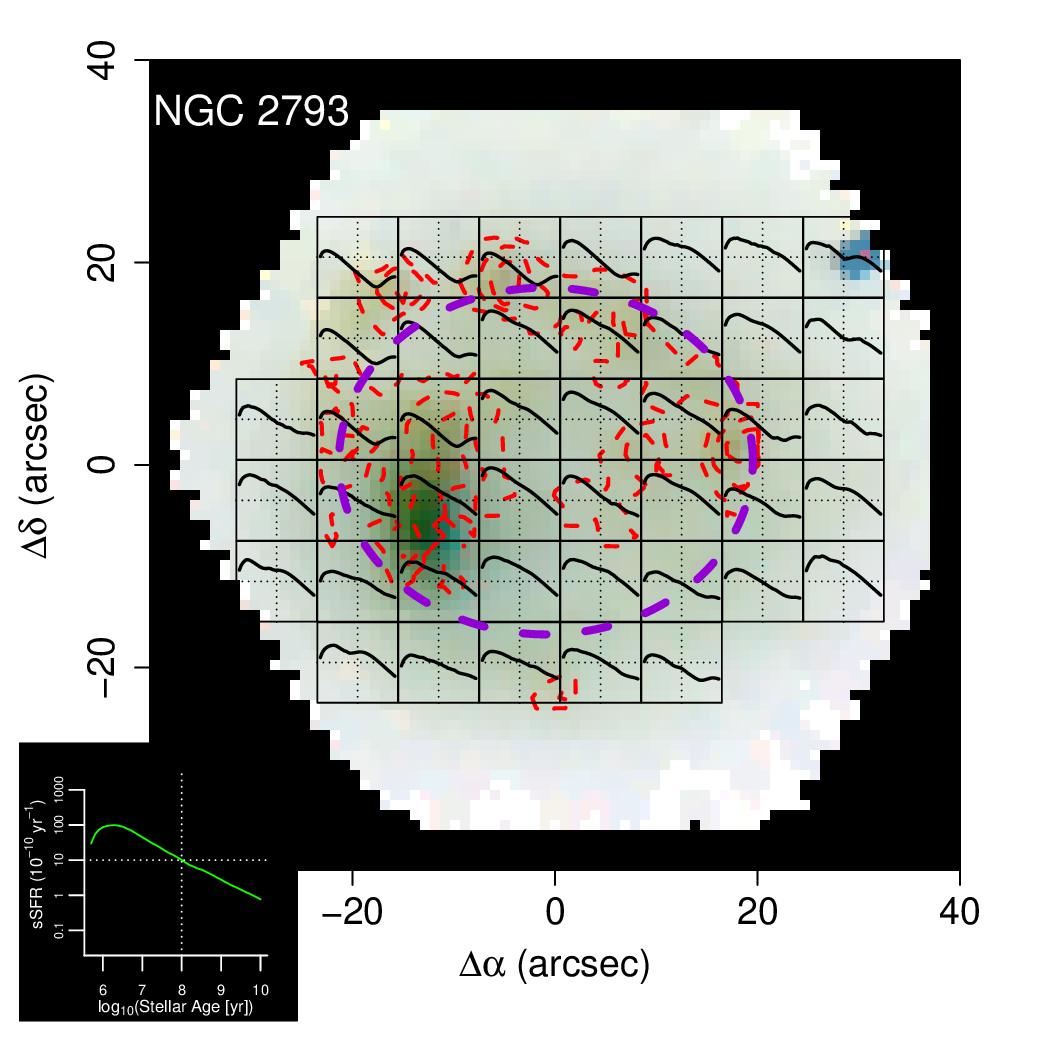}
        \caption{Same of Figure~\ref{fig:arp143_ssfh} for NGC~2793. It is important to figure out that scale is different.}
\label{fig:ngc2793_ssfh}
\end{figure*}

\begin{figure*}
  \includegraphics[width=0.95\textwidth]{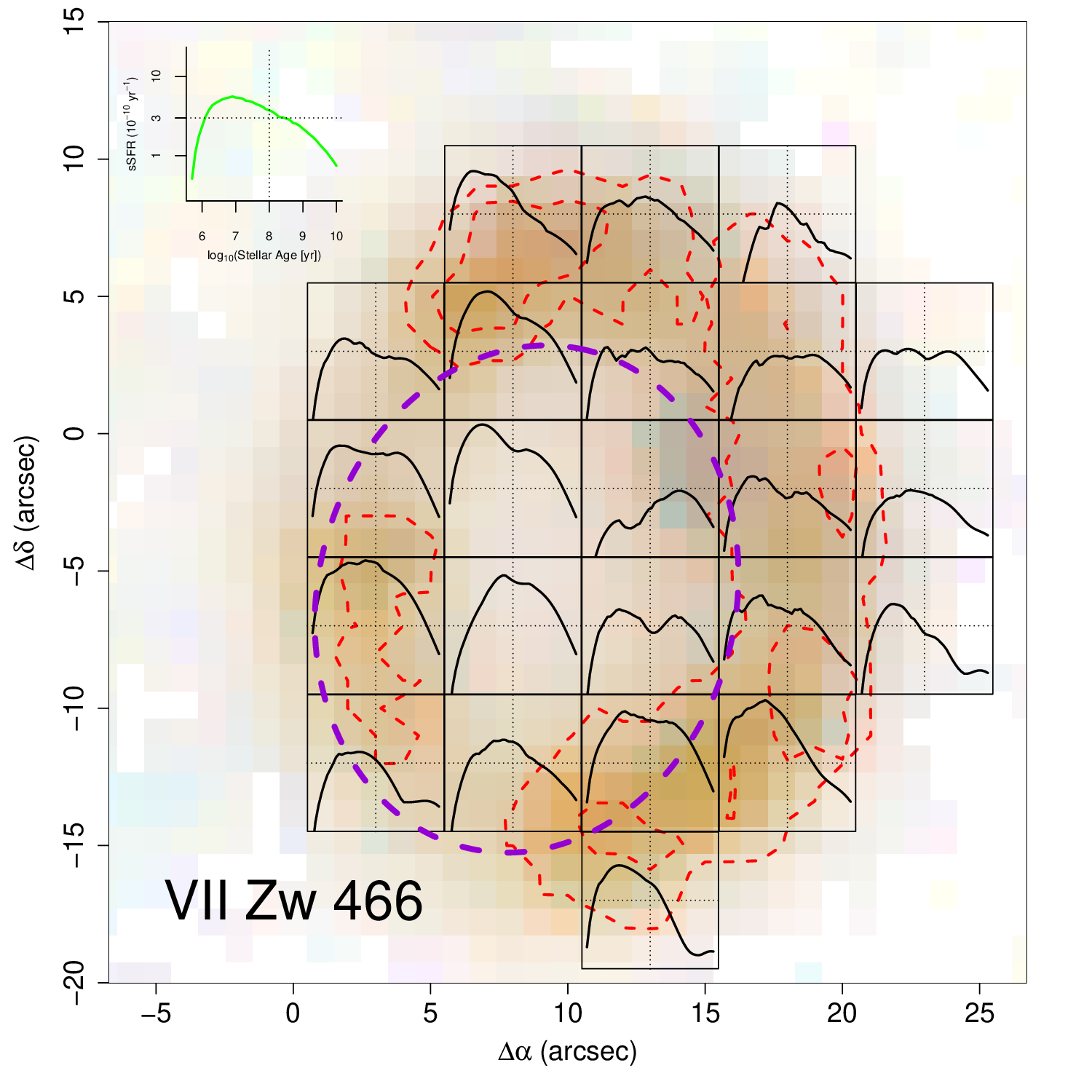}
        \caption{Same of Figure~\ref{fig:arp143_ssfh} for VII~Zw~466. Now the number of contributing spaxels was reduced to a 5$\times$5 array to compensate for the smaller angular size of this galaxy.}
\label{fig:viizw466_ssfh}
\end{figure*}

\subsection{Spatially-resolved specific star formation rates}
The IFS data allows us to carry out a spatially-resolved analysis of the sSFH (i.e., sSFR vs. stellar age, see Section~\ref{sec:viizw466}). Such maps could help us to analyse the---up to one order of magnitude---age differences for the youngest populations between the four quadrants of our classical ring galaxies Arp~143 and VII~Zw~466 (see Figure~\ref{fig:age_minboxprof}), despite all quadrants presenting similar steep negative  gradients in their youngest age profiles. As discussed before, these negative gradients are due to star formation radially propagating outwards over the last $\sim\,\,10^8$  years and are expected in ring galaxies.  

Both galaxies have ${\nabla t_\star}_y<\,\,-1.0$ dex. The ``traditional" collision hypothesis would explain these gradients as follows: the passage of the density wave through a region produces a local starburst and the light-weighted ages are dominated by the young massive stars produced in the process. As the wave abandons the region and the starburst stops, the most massive stars die and the light-weighted ages are progressively dominated by a mix of the young low mass stars generated by the wave and the old low-mass stars formed  before the collision. Therefore  the ring wave naturally produces a inside-out quenching of star formation in its wake and a light-weighted mean-age gradient is formed between the location of the ring dominated by the youngest ages and the centre (or the impact point) dominated by older ages.

In order to understand the differences in ages of the most recent star formation activity between the four quadrants, we mapped the sSFH of each of our sample galaxies. We consider the sSFR as a proxy to the SFR, and use these maps to understand the morphology of post-collision star formation.
We divided each image into an array of sectors formed by 8$\times$8 spaxels for Arp~143 and NGC~2793 to increase the S/N ratio. For VII~Zw~466, we used a smaller array of 5$\times$5 due to the smaller angular size of that galaxy. The grouping compensates for the lower S/N spaxels, allowing for a more robust coverage of the disk. The resulting median sSFH for each sector is plotted in boxes on (inverted colour) RGB maps (Figures~\ref{fig:arp143_ssfh}, \ref{fig:ngc2793_ssfh} and \ref{fig:viizw466_ssfh} for Arp~143, NGC2793 and VII~Zw~466, respectively). The median sSFH for each galaxy is included as an inset for reference. The positions of the  dotted vertical and horizontal lines are the same in all subplots as in the insets.

As the density wave created by the collision expands, it is expected to locally increase the sSFR in the sectors of the disk at which it arrives.The median sSFH of the whole galaxy for Arp~143 (bottom-left inset in Figure~\ref{fig:arp143_ssfh}) best illustrates the effect of expanding wave. It shows two main peaks of high star formation. The peak corresponding to older ages ($>10^9$ yr) represents clearly the stellar populations of the pre-collision (bulge-dominated) galaxy, with the high sSFR at the beginning of galaxy formation followed by a decrease. However, except for very well-bound regions like the bulge, this ``first'' peak is generally not representative of the regions where these stars now reside. This is because stars formed in such early star-forming episodes are unlikely to remain in their original regions due the disk rotation or other kinds of migration, (e.g., the one induced by the collision). We refer to the first peak as the ``pre-collisional peak'' and the more recent maximum as the ``post-collisional peak''. The monotonic decrease is halted at around $\log_{10}(Age \left[{\rm yr}\right])$~=~8.5, the estimated epoch of the ring-making collision. The sSFR smoothly increases then onwards. 

The sSFH in most of the sectors at the position of the ring--- especially at the H$\alpha$-emitting regions (red contours)---follow the global trend (the median sSFH of the galaxy shows in the inset in the figure), with the peak at the current epochs ($\log_{10}(Age \left[{\rm yr}\right])<$ 7.0) being stronger than the sSFR corresponding to populations of the pre-collisional peak ($\log_{10}({\rm Stellar\ Age}\left[{\rm yr}\right])>$ 9.5). 

The age when the minimum in the sSFH occurs varies only slightly in the zones with and without H$\alpha$. This suggests that even the zones in the ring, far from the impact point, contain stellar populations formed just after the collision. In the classical density wave models of \citet{theys1976}, this would not be possible, since the stellar populations formed during the passage of the wave are left behind at their formation sites. On the other hand, the stellar populations are dragged by the wave in the recent simulations by \citet{renaud2018}, allowing to find the first stellar populations formed by the impact as far out as the ring position. This scenario is consistent with our observations.

The sSFH show a distinct shape in the southern part of Arp~143, suggesting an asymmetric expansion of the ring. Some enhancement in the sSFR over the last $\sim\,\,10^8$~yr is seen, but only mildly in comparison with the northern regions. It is worth noting that the northern regions are currently closer to the companion galaxy than the southern regions. The boxes belonging to the bulge and the surrounding zones also show minimum enhancements following the collision.

Figure \ref{fig:arp143_ssfh}, also plots some boxes to the north and outside of the ring. There is a tidal bridge of stars connecting both galaxies, also visible in Figure \ref{fig:age_map}. Their sSFHs and the young values of the mean light-weighted ages are consistent with the presence of young massive stars formed as a consequence of the impact. A detailed kinematics study of this region, in synergy with the results presented here,  could reveal if these stars were formed in situ inside this bridge from gas dragged from the disk by tidal forces, or if they were formed first in the disk---most possibly in the ring---and were tidally dragged out afterwards.

From the earlier discussion, NGC~2793 does not appear to follow the collision scenario, both for the ring and the inner regions. Indeed, according to Figure~\ref{fig:ngc2793_ssfh}, it suffered recent intense star formation across the whole galaxy, with the oldest peak of sSFH not distinguishable in  most boxes. This is consistent with the flat youngest age profile shown in Figure~\ref{fig:com_age_min} in comparison to the other two galaxies (and the almost zero ${\nabla t_\star}_y$ in Table \ref{tab:gradients}). Therefore, this evidence appear to confirm that NGC~2793 is not a collisional ring galaxy. 

For the empty-ring galaxy VII Zw 466, the two peaks in the sSFR are not easily discernible  (Figure~\ref{fig:viizw466_ssfh}). The first peak---corresponding to formation of the pre-collision stellar populations---can be noticed only in a few boxes to the south and south-west outside the ring. Many of the rest of the boxes show abrupt enhancement of sSFR at $\sim 10^8$~yr, the estimated collision epoch. Remarkably, this enhancement in the star-formation activity is not just restricted to the ring regions, but is also present in the empty inner regions. The time at which the enhancement occurred does not have a clear radial dependence, suggesting that the ring contains stellar populations dragged by the wave, similar to the behaviour found in Arp~143.

\begin{figure*}
  \includegraphics[width=0.99\textwidth]{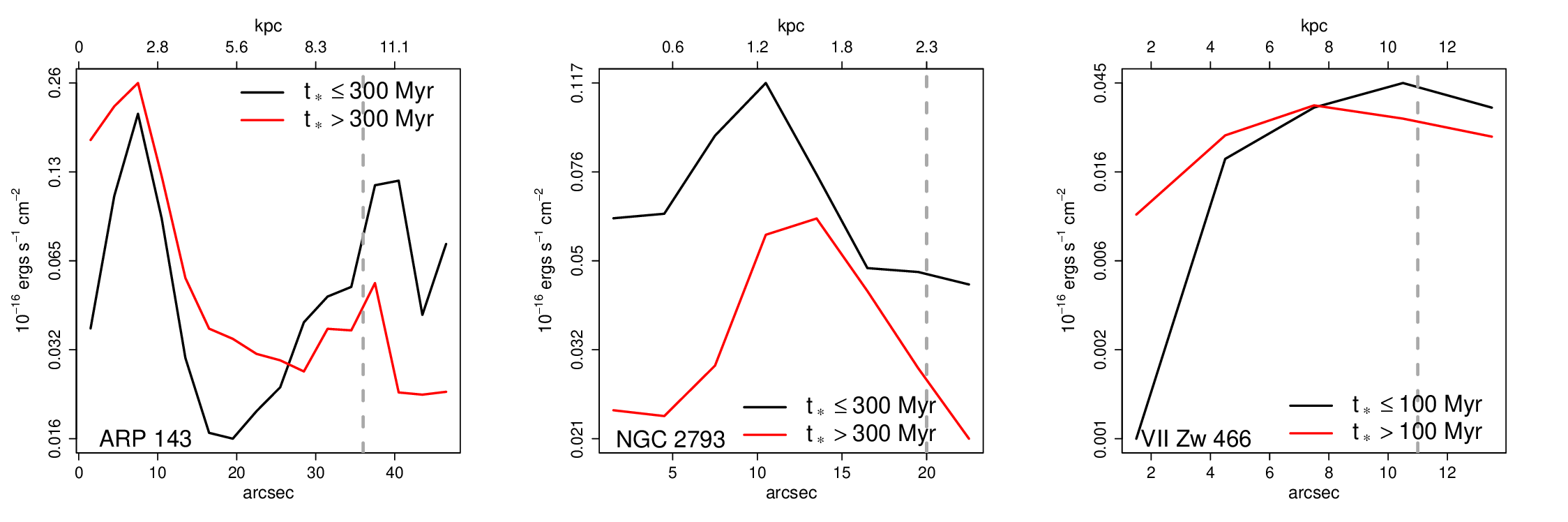}
        \caption{Comparison of the intensity profiles derived of the integrated flux of population with pre-collisional (red) and post-collisional (black) ages in the three galaxies. The current position of the ring is shown by the dotted vertical line. See the description of results for each galaxy in the text for details.
       }
\label{fig:com_age_young_and_old}
\end{figure*}

\subsection{Current location of the density wave}

The decomposition of light in stellar populations of distinct ages at each spaxel allows us to construct the radial profile of intensities from stars formed before the collision and after the collision. To do that, we separated the flux contribution of populations with pre-collision ages from those with post-collision ages in each spaxel. The flux (dust-corrected) was integrated in the range of 3800 \AA \ to 7200 \AA, i.e., the same range used for the stellar population synthesis. We then produced radial profiles that are shown in Figure~\ref{fig:com_age_young_and_old}, keeping NGC~2793 for the sake of completeness. 

According to simulations \citep{renaud2018}, the expanding density wave not only triggers new star formation in the ring but may also transport pre-existing disk stars. Consequently, the light distribution of these pre-collisional stars might  also roughly trace the density wave. Indeed, Figure \ref{fig:com_age_young_and_old} shows that while the post-collisional young populations(black solid line) dominate the optical emission at the ring radius (grey dashed line) and the old pre-collisional populations (red solid line) also shows an enhancement in flux at around that radius.

For Arp~143 (left panel  in Figure \ref{fig:com_age_young_and_old}) the peak at radius $< 10$ arcsec corresponds to light from the bulge, with contribution from both pre-collisional and post-collisional stellar populations, with the flux dominated by the former. The peak in the black line beyond the central bulge is consistent with the position of the ring and illustrates the current position of expanding density wave dragging pre-collisional stars. The figure shows that this second flux peak of pre-collisional stars is consistent with a flux peak of post-collisional stars formed by the passage of the wave. This spatial coincidence indicates that the density wave (as traced by pre-collisional stars) is effectively the most suitable cause for the presence of a star formation in the ring (as traced by post-collisional stars and H$\alpha$ emission). The fact that the young populations profile enhancement extends well beyond the ring radius while the older populations peak decays faster after that radius, can be explained by the presence of the tidal bridge in the north-western part of the ring dominated by young light-weighted ages.

For VII~Zw~466, we can see a similar pattern as with Arp~143, with the pre-collisional stars profile growing towards the ring along with the post-collisional profile, although with notable differences. First, since this galaxy is an ``empty" ring, with the possible bulge located in the inward part of the ring (therefore not close to the ring centre), the low-radius peak is not present. Second, there is a clear misalignment of both peaks, with the post-collisional profile peaking at the ring position and the pre-collisional profile peaking $\sim 3$ arcsec inward. This is explained by the presence of the bulge, strongly dominated by pre-collisional stars (like in Arp~143). Beyond this point the flux in the ring is clearly dominated by post-collisional stars (also like in Arp~143). Notice also that due to the small angular size of VII~Zw~466, spatial resolution is poorer for this galaxy and the shape of the profiles is not clearly defined. Yet the effects described above are clearly discernible.

Figure \ref{fig:com_age_young_and_old} shows that the density wave is, as expected, currently at the position of the star-formation ring in both galaxies. IFS data allowed us, for the first time, to isolate the light of pre-collisional stars from that of the post-collisional stars thus addressing the long-standing question on the current location of the density wave \citep[see for example][]{marston1995,gerber1996}. The results are in agreement with \citet{ditrani2024}, which found that most of the spectral luminosity at the ring location is dominated by such post-collisional stars for the Carthwheel galaxy.

Our data  show a density wave in the old stars, illustrating the response of these pre-collisional stars to the ring-making collision. The limited spatial-resolution and the presence of tidal features in our data prevent us from examining if the density wave is effectively ahead of the star-forming wave as expected in classical models \citep[e.g][]{gerber1996}. Future IFS observations of these  classic ring galaxies at higher spatial resolution will be invaluable in addressing this question.

\section{Summary and conclusions}

We tested the collision hypothesis of ring galaxies through a stellar population analysis with integral field spectroscopy data of three ring galaxy candidates. The galaxies come from the \citet{romano2008} sample and they were observed with the Calar Alto 3.5 m telescope, using the PMAS/PPaK spectrophotometer.

For each galaxy we produced an RGB map, the Mean Age maps weighted by light and by mass (Figure \ref{fig:age_map}), we found that light-weighted mean age  is the most appropriate to characterize the young populations created by the impact and hence the ring.

The collision hypothesis predicts the presence of a negative radial age gradient generated by the moving front of star formation triggered, which in itself is generated by the passage of the density wave. We looked for this gradient by plotting radial profiles of light-weighted mean ages. We also plotted radial profiles for the different quadrants of each galaxy account for possible azimuthal asymmetries (Figure \ref{fig:age_boxprof}).

Then we produced another set of radial profiles that represent better the last major episode of star formation at each radius by statistically selecting the youngest spaxels (in terms of light-weighted mean age) at each radius (\ref{fig:age_minboxprof}).  This allowed us to compute the youngest ages gradient (${\nabla t_\star}_y$) as a function its distance from the ring centre. With these gradients we estimated the lower limits for the age of the collision and for the radial velocity of the wave. These calculated quantities are presented in Table~\ref{tab:result}. An sSFH analysis was also performed using Equation~\ref{eq:ssfh}. 

We found that the Arp~143 and VII~Zw~466 galaxies show negative age gradients consistent with the collision hypothesis, with the youngest population ageing  smoothly as we move from the ring position inwards. We estimated a lower limit for the collision age of 300 Myr and 100 Myr for Arp~143 and VII~Zw~466, respectively, assuming that the youngest ages at the ring centres correspond with the collision age. Assuming a uniform propagation, this corresponds to average  expansion velocities of 33\,\,$\pm$\,\,10 and 108\,\,$\pm$\,\,28 km s$^{-1}$, respectively for Arp~143 and VII~Zw~466 (Table~\ref{tab:result}). 

An analysis of the population properties in the four quadrants suggests noticeable azimuthal variations in the ring which could point off-centre geometries for the collision. We created spatially-resolved specific star formation history (sSFH) maps, by dividing the galaxies in boxes containing a given number of spaxels to increase the SNR. These maps show considerable azimuthal variations with some regions not showing any enhancements in star formation corresponding to the propagation of the wave. In zones where enhancements in sSFRs are noticeable, the regions in the ring as well as those near the impact point, show enhancements at the same time. This suggests that the stellar populations formed during the propagation of the wave could have been dragged by the wave to its current position in the star-forming ring, as predicted in the recent models for Cartwheel-like rings by \citet{renaud2018}. 

We also measured the redshifts of the four members of the VII~Zw~466 group, confirming that one of them  (B1) is really a background galaxy. We used the sSFH of the other three galaxies, which showed that G2 has recent star formation activity consistent with the one in the ring, which is not the case with the other group member G1. Therefore, G2 was most likely the galaxy that impacted with VII Zw 466 causing the formation of the ring. This is consistent with the HI bridge connecting both galaxies reported by \citet{appleton1996}, and confirming their results.

In comparison, the behaviour of the stellar population ages in NGC~2793 is different from the other two galaxies. We notice that this is a low-mass galaxy and one of the smallest rings of the \cite{romano2008} sample, with only $\sim 2$~kpc radius. Star formation is expected to be quenched as the wave moves outwards due to negative feedback effects of supernova explosions and gas consumption over a timescales a few tens of million years. The recent star formation is very similar on all the regions of the galaxy, both in the ring and in the inner regions. In addition, the age of the supposed ring is about $10^7$~years~old  Table~\ref{tab:result}), but its value is in the range of the young populations across the whole galaxy and not only in the ring. All this evidence support that NGC~2793 is no a true collisional ring galaxy. 

In this work, we showed that the combination of IFS and stellar population synthesis is  an invaluable tool for observationally testing the predictions of different theoretical collisional scenarios. The application of these techniques to larger samples of collisional ring galaxy candidates combined with higher spatial and spectral resolution instruments will help to disentangle the details and history of such extreme collisions and the transformation suffered by galaxies that pass through this process.

\section*{Acknowledgements} 
This paper is proudly, to our knowledge, the first ever peer-reviewed paper in astrophysics lead by astronomers working from Nicaragua. M. C.-M. and A. R.-O. want to dedicate this work to José~ H.~Peña from the Instituto de Astronomía at UNAM (Mexico). His long-standing efforts to bring formation in true professional astronomy to the youth in Central and South America ultimately made this achievement possible.

We thank the referee for the insightful comments that helped to improve the quality of this paper.

R. G.-B. acknowledges financial support from the Severo Ochoa grant CEX2021-001131-S funded by MCIN/AEI/10.13039/501100011033 and to grant PID2022-141755NB-100.

\section*{Data availability}
The observational data are publicly available through the Calar Alto Observatory archive (https://www.caha.es/access-and-services/public-archives) under program
F17-3.5-027 with PI R. Ortega-Minakata. Programming code used for this analysis will be made available on reasonable request to the corresponding author. 



\bibliographystyle{mnras}
\bibliography{mnras} 








\bsp	
\label{lastpage}
\end{document}